\documentstyle[aps,twocolumn,epsfig,rotating]{revtex}

\begin{document}




\begin{title} {\bf Muon spin relaxation studies of incommensurate  
magnetism and superconductivity in stage-4 La$_{2}$CuO$_{4.11}$
and La$_{1.88}$Sr$_{0.12}$CuO$_{4}$\/}
\end{title}

\author{A.T.~Savici,  
Y.~Fudamoto, I.M.~Gat, T.~Ito, M.I.~Larkin, Y.J.~Uemura $^{1}$
}

\address{Department of Physics, Columbia University, 
New York, NY 10027, USA}

\author{G.M.~Luke}

\address{Department of Physics and Astronomy, McMaster University,
Hamilton, Ontario L8P 4N3, Canada}

\author{K.M.~Kojima}

\address{Department of Superconductivity, Faculty of Engineering,
University of Tokyo, 7-3-1 Hongo, Tokyo 113-8656, Japan}

\author{Y.S.~Lee, M.A.~Kastner}

\address{Department of Physics, MIT, Cambridge, MA 02139, USA}

\author{R.J. Birgeneau}

\address{Department of Physics, MIT, Cambridge, MA 02139, USA, and}

\address{Department of Physics, Univ. of Toronto, Toronto,
Ontario M5S 1A7, Canada}

\author{K.~Yamada}

\address{Institute for Chemical Research, Kyoto University,
Uji, Kyoto 611-0011, Japan}

\vspace{-5truemm}

\date{\today}

\maketitle



\begin{abstract}
We report muon spin relaxation ($\mu$SR) measurements
using single crystals of oxygen-intercalated stage-4 La$_{2}$CuO$_{4.11}$ 
(LCO:4.11) and La$_{1.88}$Sr$_{0.12}$CuO$_{4}$ (LSCO:0.12), in which 
neutron scattering studies have found incommensurate magnetic 
Bragg reflections.
In both systems, zero-field $\mu$SR measurements 
show muon spin precession below the N\'eel temperature $T_{N}$  
with frequency 3.6 MHz at $T\rightarrow$ 0, having
a Bessel function line shape, characteristic of spin-density-wave systems.  
The amplitude of the oscillating and relaxing signals 
of these systems is less than half the value expected for
systems with static magnetic order in 100 \%\ of the volume.
Our results are consistent with a simulation of local fields
for a heuristic model with: (a) incommensurate
spin amplitude modulation with the maximum ordered Cu moment size 
of $\sim$ 0.36 $\mu_{B}$; 
(b) static Cu moments on the CuO$_{2}$ planes forming ``islands'' having 
typical radius $15\sim 30$ \AA, comparable to the in-plane 
superconducting coherence length;
and (c) the measured volume fraction of magnetic muon sites $V_{\mu}$ 
increasing progressively
with decreasing temperature below $T_{N}$ towards $V_{\mu} \sim 40$ \%\ 
for LCO:4.11 and 18 \%\ for LSCO:0.12 at $T\rightarrow 0$.
These results may be compared with correlation lengths in excess 
of 600 \AA\ and a long range ordered moment of $0.15 \pm 0.05$ $\mu_B$ 
measured with neutron scattering techniques.  
In this paper we discuss a model that reconciles these apparently 
contradictory results.
In transverse magnetic field $\mu$SR measurements, 
sensitive to the in-plane magnetic 
field penetration depth $\lambda_{ab}$, the results 
for LCO:4.11 and LSCO:0.12 follow 
correlations found for underdoped, overdoped and Zn-doped high-$T_{c}$
cuprate 
systems in a plot of $T_{c}$ versus the superconducting relaxation rate
$\sigma(T\rightarrow 0)$.  
This indicates that the volume-integrated
value of $n_{s}/m^{*}$ (superconducting carrier density / effective
mass) is a determining factor for $T_{c}$, not only in high-$T_{c}$
cuprate systems
without static magnetism,
but also in the present systems where superconductivity co-exists with 
static spin-density-wave spin order.
\footnote{author to whom correspondences should be addressed.
E-mail: tomo@lorentz.phys.columbia.edu}
\end{abstract}


%



\narrowtext
\section{Introduction}
The interplay between superconductivity and magnetism is one of the central
issues of high-$T_{c}$ superconductivity (HTSC), which has been
extensively studied both experimentally and theoretically.[1].  
In particular, dynamic [2-8] and
static [9-16] spin correlations with incommensurate wave vectors have been
found by neutron scattering.  These, as well as 
X-ray photoelectron studies [17-19], have been discussed in 
terms of a ``stripe'' 
modulation of the spin and charge densities [20-23].
Yet it has not been clear whether static magnetism 
supports or competes with superconductivity, or if the magnetism
and superconductivity coexist
in the same microscopic regions of the CuO$_{2}$ planes or in different
regions.  While neutron measurements make it clear that 
long range magnetic order exists, the
Bragg peak intensity integrates over the sample volume and is not, 
therefore, sensitive to microscopic spatial variations in the order parameter.

Muon spin relaxation ($\mu$SR) measurements [24,25] provide a complementary
probe in this regard.  In a magnetic material having a heterogenous
structure, such as ``magnetic'' and ``non-magnetic'' regions,
$\mu$SR data are composed of two different signals, corresponding
to different environments with signal amplitudes roughly
proportional to their volume fractions.  Local 
magnetic fields at muon sites result primarily from the dipolar interaction.  
In antiferromagnetically ordered systems the local field decays very quickly
with increasing distance from ordered spins, resulting in an 
effective range of the field of $\sim$ 10-15 \AA\ in cuprate systems.  
Thus, any heterogeneous magnetic structure in HTSC, having
a length scale larger than this, should produce multiple
and distinguishable $\mu$SR signals.

In this paper, we report zero-field (ZF) and transverse-field (TF)
$\mu$SR measurements in a single crystal of La$_{2}$CuO$_{4.11}$ (LCO:4.11)
where excess oxygen is intercalated in a stage-4 structure [16].   
This system is superconducting below
$T_{c} \sim 42$~K.  Observation of sharp satellite Bragg peaks 
in neutron scattering indicates static long range ($>$ 600 \AA) 
spin density wave (SDW) order below 
$T_{N} \sim 42$~K.
The modulation wave vector of the SDW is comparable to those observed in
La$_{2-x}$Sr$_{x}$CuO$_{4}$ (LSCO) and La$_{2-x-y}$Nd$_{y}$Sr$_{x}$CuO$_{4}$
(LNSCO) systems with the hole concentration $x \sim$ 0.125 [9,12]. 
The LCO:4.11 system is especially interesting since
(a) it has the highest superconducting $T_{c}$ in the 214 
family of materials;
(b) it has a rather high magnetic
$T_{N}$, which is very close to $T_{c}$, and well developed
long range magnetic order below $T_{N}$;
(c) there is no randomness arising from Sr substitution, and the 
intercalated oxygen ions are
three dimensionally ordered.
Preliminary results and analysis of our ZF-$\mu$SR in LCO:4.11 have been
presented at a recent conference [26].

We also report $\mu$SR results 
in a superconducting La$_{1.88}$Sr$_{0.12}$CuO$_{4}$ (LSCO:0.12) 
single crystal 
with $T_{c} \sim 30$ K which exhibits magnetic Bragg peaks in 
neutron scattering [13].  
Comparing these results
with those of a simulation of the local field distribution at 
possible muon sites,
we examine various models for the magnetically ordered regions.
The interplay between superconductivity and magnetism is
discussed in the context of the superfluid density $n_{s}/m^{*}$ derived from 
TF-$\mu$SR results.

$\mu$SR studies of magnetism in HTSC systems started in 1987 with
studies of antiferromagnetic La$_{2}$CuO$_{4}$ (AF-LCO) [27], which provided
the first evidence of static magnetic order.  Subsequently,
$\mu$SR studies on La$_{2}$CuO$_{4}$ [28], 
(La,Li)$_{2}$CuO$_{4}$ [29], La$_{2-x}$Sr$_{x}$CuO$_{4}$ [30-34],
and YBa$_{2}$Cu$_{3}$O$_{y}$ [34,35], were performed to characterize
magnetic phase diagrams as a function of hole 
concentration and also to study details of the spin glass
states near $x = 0.05$.  

$\mu$SR measurements of HTSC systems
with hole concentration $x$ near 1/8 were 
first performed in La$_{1.875}$Ba$_{0.125}$CuO$_{4}$ [36].  
These detected static magnetic order below $T_{N}$ = 32 K 
and a Bessel function $\mu$SR line shape, characteristic of spin density wave
systems.  A reduction of $T_{c}$, associated with an increase of the muon 
spin relaxation due to quasi-static magnetism, was also found by
$\mu$SR in La$_{2-x}$Sr$_{x}$CuO$_{4}$ near $x =0.12$ [37].
Recent $\mu$SR studies on 
(La,Nd,Sr)$_{2}$CuO$_{4}$ (LNSCO) [38,39],  
(La,Eu,Sr)$_{2}$CuO$_{4}$ (LESCO) [40,41] and
La$_{1.875}$Ba$_{0.125-y}$Sr$_{y}$CuO$_{4}$ (LBSCO) [42]
found the characteristic 
Bessel function line shape with the same frequency $\nu \sim 3.5$ MHz
in all these systems.  In the latter studies the primary emphasis 
was determining the
magnetic phase diagram as a function of rare-earth and hole
concentrations.  Although the existence of ``zero/low field muon 
sites'', expected for decomposition of the system into regions with and without 
static magnetic order, was discussed in these reports [40,42],
a systematic study of magnetic volume fraction was not made.

Previously, Pomjakushin {\it et al.\/} [43] 
reported a $\mu$SR study for
single crystal specimens of La$_{2}$CuO$_{4+y}$ doped with oxygen,
with $y$ = 0.02 (LCO:4.02) and 0.04 (LCO:4.04), which have 
compositions in the miscibility
gap between AF-LCO and the stage-6 superconducting compound.  
In zero field, they observed $\mu$SR precession signals identical to 
those found in AF-LCO.  With decreasing temperature in LCO:4.02, 
the amplitude of this oscillatory signal increased from nearly zero
above the superconducting $T_{c} \sim 15$ K to more than a half of
that in AF-LCO at $T\rightarrow 0$.  In LCO:4.04,
the oscillatory signal appeared below $T\sim 230$ K
with an amplitude corresponding to approximately a 10 \%\ volume
fraction for the AF region, and then exhibited a sharp increase
below the superconducting $T_{c} \sim 25$ K to about a half volume 
fraction at $T\rightarrow 0$.  In both cases, the amplitudes of the
AF oscillation exhibited an increase below $T_{c}$. 
The diamagnetic susceptibility of the LCO:4.02 specimen was, however,
destroyed by a small applied field, indicating rather 
fragile superconductivity.  
Lack of information about the crystal orientation in the $\mu$SR 
measurements prevented a reliable estimate
of the volume fraction in this study.

In Section II we present experimental details and results of 
the ZF-$\mu$SR measurements.  
Section III contains our simulation based on a model in 
which the sample contains microscopic regions where there 
is full magnetic order and other microscopic regions
where the magnetic fields are too small to cause muon 
precession.  In Section IV we presents results of measurements in
a transverse magnetic field that are sensitive to the 
superconductivity.  Finally in Section V we discuss the results and
draw conclusions.

\section{Zero field $\mu$SR measurements}

Single crystals of LCO:4.11 and LSCO:0.12 have been prepared as
described in refs. [16] and [13],
which report neutron scattering studies of these crystals, respectively.
The excess-oxygen doped La$_2$CuO$_{4+y}$ sample is prepared by
electrochemically doping a single crystal of pure La$_2$CuO$_4$
which is grown by the traveling solvent floating-zone method.
The crystal has a mass of 4.21 grams and is cylindrical in
shape.  As result of twinning, there are equal populations of
two twin domains with either the a or b crystallographic axis
(in orthorhombic notation) nearly parallel to the cylinder's
long-axis.  In both twin domains, the c-axis is perpendicular to
the long-axis.  
Thermogravitimetric analysis on our samples has been performed
on two smaller La$_2$CuO$_{4+y}$ crystals with $T_c = 42$K.  We
find oxygen concentrations of $y=0.10(1)$ and $y=0.12(1)$.
The large crystal has the same $T_c$, so its oxygen concentration 
is expected to correspond to $y=0.11(1)$.

Neutron diffraction measurements have shown that this
material has a structural modulation along
the c-axis corresponding to stage-4.
[16].  Further details regarding the
characterization of the stage-4 crystal by
neutron scattering, susceptibility, and transport techniques may
be found in references [16] and
[44].

$\mu$SR measurements
were performed at TRIUMF, 
using a surface muon beam with an incident muon momentum of 28 MeV/c,
following the standard procedure
described in refs. [24,25].  
We employed the so-called ``low-background'' muon spectrometer,
which has the capability of vetoing events from muons 
landing in areas other than the specimen, to ensure much less than $\sim$ 
10 \%\ background signal in our $\mu$SR data.
In ZF-$\mu$SR, we observed positron time spectra
via counters placed forward (F) and backward (B) to the incident beam
direction ($\hat z$), with the polarization of incident muons $\vec{P}_{\mu}$
anti-parallel to $\hat{z}$.  
The muon spin relaxation function $G_{z}(t)$ was obtained
from forward (F) and backward (B) time spectra as
$$G_{z}(t) = [B(t) - F(t)]/[F(t) + B(t)] \eqno{(1)},$$
after correcting for the difference in effective solid angles of the
F and B counters. As illustrated in Fig. 1(a),
two sets of measurements were made with the
c-axis of the specimen mounted parallel (configuration CZ-I)
and perpendicular (CZ-II) to $\hat{z} \parallel \vec{P}_{\mu}$.

Figures 2(a) and (b) show the muon spin polarization function $G_{z}(t)$
observed in these two crystals for CZ-I and CZ-II.
Also included are the results obtained for ceramic specimens of 
non-superconducting La$_{1.875}$Ba$_{0.125}$CuO$_{4}$ (LBCO:0.125) [36]
and La$_{1.475}$Nd$_{0.4}$Sr$_{0.13}$CuO$_{4}$
(LNSCO:0.13) [38]. 

We see almost no relaxation above $T_{N}$ in the time range $t \leq$ 
1 $\mu$s.  Below $T_{N} \sim 42$ K of LCO:4.11 and $T_{N} \sim 20$ K
of LSCO:0.12, a small oscillatory component appears, with a Bessel-function
line shape characteristic of muon precession in SDW systems [45,46].
The amplitude $A_{or}$ of this oscillating and relaxing 
signal is limited to about 34 \%\
(CZ-I) and 24 \%\ (CZ-II) at $T\rightarrow 0$ for LCO:4.11, and
20 \%\ (CZ-I) and 10 \%\ (CZ-II) for LSCO:0.12.  
If we define the internal field at a muon site as $\vec{H}$ 
and $\theta$ is the angle between $\hat{z}$ and $\vec{H}$,
as illustrated in Fig. 1(b),
the oscillatory / relaxing amplitude represents $\sin^{2}(\theta)$
averaged over all the muon sites.  In a ceramic specimen
which undergoes magnetic order in 100 \%\ of the volume, this 
amplitude is 67 \%\ (=2/3), as shown in the case of LBCO:0.125.
Therefore, the observed signal amplitudes $A_{or}$ for LCO:4.11 and LSCO:0.12
are less than a half the value expected for a sample with a SDW uniformly 
established throughout its volume. 

In single-crystal specimens, $A_{or}$ can have a small value
if the local fields $\vec{H}$ for the majority of muon sites are parallel 
to $\hat{z} \parallel \vec{P}_{\mu}$.
For this case, however, one should see a large $A_{or}$ when
the crystal orientation is rotated by 90 degrees 
from CZ-I to CZ-II or vice versa.
Thus, our results, with small $A_{or}$ values for the
both configurations, cannot be attributed to such anisotropy.
In view of the symmetry of the LCO and LSCO structures, the 
``powder average'' value of $A_{or-pwd}$ 
can be obtained [26] as 
$$A_{or-pwd} \equiv (1/3)\times[A_{or:CZ-I} + 2A_{or:CZ-II}], \eqno{(2)},$$ 
which yields
27 \%\ for LCO:4.11 and 12 \%\ for LSCO:0.12.  
The reduction of
$A_{or-pwd}$ from 66.7 \%\ indicates the existence of 
``zero/low-field''  muon sites.

We have analysed the ZF-$\mu$SR data by fitting the results
at $t\leq 1 \mu$s to
$$G_{z}(t) = A_{or}j_{o}(\omega t)exp(-\Lambda t) +
A_{n}exp[-\Delta^{2}t^{2}/2] \eqno{(3)},$$
where $j_{o}$ is the 0-th order Bessel function and $A_{n}$ represents 
the non-oscillating amplitude. The first exponential term
represents the damping of the Bessel oscillation, whereas the second 
describes the slow decay due to nuclear dipolar fields.
We obtain the volume fraction $V_{\mu}$ of muon sites that 
experience the SDW field using 
$$V_{\mu} \equiv A_{or-pwd}/(2/3), \eqno{(4)}$$ 
with A$_{or-pwd}$ determined from Eq. (2) at the
lowest temperature.  
We then
scale the values for higher temperatures using $A_{or}$ results for the CZ-I
configuration.
For our fit range of $t \leq 1 \mu$s, those muon sites
with $H \geq 30$ G contribute to the first term in Eq. 3, and thus
are identified as ``finite-field muon sites''.

Figures 3(a) and (b) show the temperature dependence of $V_{\mu}$ and 
the oscillation frequency $\nu = \omega/2\pi$ for LCO:4.11, LSCO:0.12,
LBCO:0.125 [36] and LNSCO:0.13 (ceramic) [38]. 
The common Bessel-function line shape and the nearly identical frequency
$\nu \sim 3.5$ MHz (corresponding to maximum internal field of 260 G)
at $T\rightarrow 0$ for all the four systems indicate that 
the spin configurations
and magnitudes of the static Cu moments (proportional to $\nu$)
surrounding the ``finite-field muon sites'' are identical for all
the four systems.  
The volume fraction $V_{\mu}$ increases progressively below $T_{N}$
with decreasing temperature, in both LCO:4.11 and LSCO:0.12.
The frequency $\nu$ acquires its full value just
below $T_{N}$ in LCO:4.11.      
In Fig. 3(c), we compare the 
temperature dependence of the neutron Bragg-peak intensity $I_{B}$
of LCO:4.11 [16] with $V_{\mu} \times \nu^{2}$ from $\mu$SR.  
The good agreement of neutron and muon results indicate that these
two probes detect the same static magnetism even though the former measures
the ordered moment over distances $\geq600$ \AA.  The $\mu$SR 
results suggest that the observed temperature
dependence of $I_{B}$ should be ascribed mainly to the change 
of the site fraction containing ordered Cu spins, rather than to the increase 
of the static moment on individual Cu atoms although this is model dependent.

To illustrate further this unusual temperature dependence of
the magnetic order
parameter in LCO:4.11, we compare ZF-$\mu$SR spectra 
of LCO:4.11 and AF-LCO ($T_{N} \geq 250$ K) in Fig. 4.
In AF-LCO (Fig. 4(b)), as in most conventional 
magnetic systems where long-range magnetic order
develops in 100 \% of  the volume, ZF-$\mu$SR data exhibit muon spin precession 
whose
amplitude is nearly independent of temperature, but whose frequency 
increases with decreasing temperature. In contrast, for LCO:4.11 (Fig. 4(a)) 
the oscillation amplitude 
increases gradually with decreasing temperature below $T_{N}$ while
the frequency just below $T_{N}$ is already close to the
low-temperature value.  This implies that in LCO:4.11 the local order 
is well developed
but dynamic above $T_N$, and the ordering process below $T_N$ involves 
the ordering of this local magnetism over
long distances with a concomitant freezing of the spins.

\section{Simulation of zero-field $\mu$SR}

To explore the possible origin of the two kinds of muon sites, 
we have performed computer simulations of the local 
field distributions at possible muon sites.  The location of the muon site in 
AF-LCO has been approximated by Hitti
{\it et al.\/} [47] to be about 1 \AA\ away from the apical oxygen,
based on the expected length of the O$^{2-}$-$\mu^{+}$ hydrogen bond. 
We have improved their estimate by taking into account the tilt of the
CuO$_{6}$ octahedra and by comparing the observed $\mu$SR
precession frequency $\nu = 5.8$ MHz 
with the Cu long-range ordered moment size 0.55 $\mu_{B}$ in AF-LCO [16].
Figure 5 shows the location of the apical-oxygen muon site
at (0.199$d$, 0$d$, 0.171$c$) in the unit cell
with $d$ = 3.779 \AA, and $c$ = 13.2 \AA\  assumed
in this simulation.
Since the direction of $\vec{H}$ is very sensitive to a small change of the 
tilt angle of the CuO$_{6}$ octahedra, we calculate the expected ZF-$\mu$SR 
line shape for the powder-average.

Figure 6(a) shows the line shape for this apical-oxygen muon site
for: (1) a 100 \%\ volume fraction of antiferromagnetically (AF) ordered Cu
moments having modulation vector $k_{AF}$ with ordered moment of 
0.55 $\mu_{B}$ (chain-dotted line); and (2) Cu spins with 
an incommensurate SDW amplitude
modulation superimposed on the 
antiferromagnetic correlations,
with a maximum static Cu moment value 
of 0.35 $\mu_{B}$ (broken line). 
Here we have assumed that the SDW propagation vector is along the 
Cu-O-Cu bond direction (i.e., 45 degrees to the orthorhombic
$a$ and $b$ axes) with magnitude $k = 2\pi/d \times 0.12$
($d = 3.78$ \AA\ is the distance between nearest neighbour Cu atoms)
We have calculated the spin polarization using
$$\cos(\vec{k}_{AF}\vec{x})\times\cos(\vec{k}\vec{x})=$$
$$(1/2)\cos[(\vec{k}_{AF}+\vec{k})\vec{x}] 
+ (1/2)\cos[(\vec{k}_{AF}-\vec{k})\vec{x}], \eqno{(5)}$$
at a given position $\vec{x}$.
These results confirm that the SDW model gives
a line shape nearly identical to a Bessel function (solid line).
In AF-LCO the observed $\mu$SR results [27,47] show a substantial 
damping of the sinusoidal precession, due to various possible
origins, such as nuclear dipolar fields, variations of
the muon site or crystal imperfection.  
To account for the reduction of
the observed $\mu$SR frequency from 5.8 MHz in AF-LCO 
to 3.64 MHz at $T\rightarrow 0$ in LCO:4.11, 
we need to assume that the maximum static Cu moment size in the SDW modulation
is $\sim$ 0.36 $\mu_{B}$.  This may be compared with the bare Cu 
moment of 1.1 $\mu_{B}$.

Since there is a substantial density of intercalated oxygen in LCO:4.11,
we have also modeled the field distribution at several locations
1 \AA\  away from the expected site of the intercalated oxygen atoms.  
As may be seen from Fig. 6(b),
which displays decay curves
calculated for 100 \%\ ordered SDW Cu moments with maximum moment
size of 0.35 $\mu_{B}$,
there is no muon site location having zero or low internal field.
Therefore, we cannot ascribe the existence of ``zero/low-field sites''
in LCO:4.11 to intercalated oxygen.  Furthermore, in LSCO:0.12, 
there is, of course, no
intercalated oxygen, so it seems clear that the non-magnetic sites 
must have a different origin.
  
In Fig. 7(a), we show the line shape $G_{z}(t)$ calculated for Cu moments
in circular islands of SDW order in the
CuO$_{2}$ planes, for various volume fractions $V_{Cu}$ of 
static Cu spins, assuming the radius $R$ of the circular
islands to be $R \sim 50$ \AA.  Bessel function line shapes 
are obtained with a common frequency, while the regions without
static Cu moments account for the zero-field sites.  
The size of the magnetic island is reflected in the dampening of
the Bessel oscillation.  

In Fig. 7(b) we compare the results of simulations for two different 
geometries.  One is for circular 
magnetic islands with $V_{Cu} = 50$\%\ and
$R$ = 15 and 100 \AA.  The other corresponds to an ``ordered sandwich'' 
model, where two out of every four CuO$_{2}$ planes contain completely 
ordered SDW Cu spins while the other two planes, presumably those adjacent 
to the 
intercalated O layers, have no static Cu spins.
The curve for $R = 15$ \AA\ islands shows a fast dampening
of the Bessel oscillation,
together with a slow decay of the 
non-oscillating signal.  
In contrast, the curves for
$R = 100$ \AA\ islands and for the sandwich model exhibit a long-lived
Bessel oscillation, with little decay of the non-oscillating signal.

After generating line shapes for the SDW-island model, 
we have fitted the simulation results to the functional form of
Eq. 3.  Figure 8(a) shows the resulting magnetic site
fraction $V_{\mu}$ as a function of volume fraction containing static 
Cu spin order
$V_{Cu}$.  We see that $V_{\mu}$ increases linearly with 
increasing Cu fraction, with a slope that is independent of island size $R$.  
The site fraction $V_{\mu}$ 
becomes nearly 100 \%\ at $V_{Cu} \sim 75$\%.  
This implies that even the observation of a magnetic $\mu$SR signal 
corresponding to $V_{\mu}=100$ \%\  cannot rule out the existence of 
regions of non-ordered Cu moments in $\sim$ 25 \%\ of the volume.  
By using the relationship between $V_{\mu}$ and $V_{Cu}$,
we estimate as lower limits $V_{Cu} \sim 27$ \%\ for our crystal 
of LCO:4.11 and
$\sim 10$ \%\ for LSCO:0.12. Of course, this same possible 
difference between $V_{\mu}$ and $V_{Cu}$
applies in all other copper oxide SDW systems.

In Fig. 8(b), we show the dampening rate $\Lambda$ of the 
Bessel function oscillation.  We find that a smaller
radius $R$ of the island results in a larger 
$\Lambda$.  We also fit the observed spectra in 
LCO:4.11 with Eq. 3 and plot the corresponding point
in Fig. 8(b) (open star symbol).  The result
$\Lambda_{4.11} = 1.25 \mu s^{-1}$ 
agrees well with the simulation for $R = 15$ \AA.
In actual systems, however, additional factors 
cause dampening of the Bessel oscillation, such as 
variations in the muon site, nuclear dipolar fields, 
and the effects of imperfections.

To account for these additional effects, we have fit the spectrum of
AF-LCO at $T$ = 20 K in Fig. 4(b) 
with the simple form $\cos(\omega t + \phi)$
multiplied by an exponential decay $\exp(-\Lambda t)$,
and obtain $\Lambda_{AF} \sim 0.81
\mu s^{-1}$.  Assuming that $\Lambda_{AF}$ represents the
additional relaxation contributions, we have obtained a corrected
relaxation rate from
$$\Lambda_{4.11}^{c} \equiv (\Lambda_{4.11}^{2} - \Lambda_{AF}^{2})^{1/2}.
\eqno{(6)}$$ 
That is, we assume that  the decay constants add in quadrature.

$\Lambda_{4.11}^{c}$ is shown by the closed star symbol in Fig. 8(b). 
These considerations lead us to
estimate $R = 15 \sim 30$ \AA\
for the average radius of the static magnetic islands
in LCO:4.11.   A similar analysis for LSCO:0.12
would give roughly the same estimate for $R$.  
However, because the oscillation amplitude in LSCO:0.12 is much smaller than
that in LCO:4.11, the analysis would be less reliable.

It is also possible to obtain the relaxation rate expected for 
the sandwich model.  The result ($\Lambda = 0.2 \mu s^{-1}$), 
shown in Fig. 8(b) by the
asterisk symbol, indicates that this model is less successful 
than the magnetic island model
in reproducing the relaxation $\Lambda_{4.11}^{c}$ observed in 
LCO:4.11.

We now discuss a simple heuristic model for the propagation of magnetic order
between these presumed islands.
If we assume that spins in neighbouring islands in the same 
CuO$_{2}$ plane are correlated when there is any overlap in the islands'
area, 
as illustrated in 
Fig. 9, we can estimate the effective correlation length of the spin order.  
For various values of $V_{Cu}$
we have calculated the probability that the direction of the Cu spins
in the magnetic islands, located at a distance $r$ from a Cu spin
in the center of a magnetic island, is correlated with the direction of
the Cu spin at $r=0$. The results of this 
calculation, shown in Fig. 9,
suggest that even with the assumption that the spins are in islands, 
the order in a given plane may nevertheless propagate over long distances.  
However, for V$_{Cu} = 30$\% the correlation length from Fig. 9 is 
only $\sim3$ times the size of the magnetic
island, whereas the narrow SDW peaks observed in neutron scattering
measurements indicate that the static spins are correlated over
very large distances ($>$ 600 \AA) within the plane [16].
Therefore, some additional mechanism to increase the
connectivity is required.

The results in Fig. 9 are obtained assuming no spin correlations
between magnetic islands belonging to different neighbouring CuO$_{2}$ planes.
The neutron measurements of LCO:4.11 [16] indicate that the static 
SDW order exhibits short range correlations along the c-axis direction with 
a correlation length of $\sim 13$ \AA, somewhat larger than 
twice the distance between adjacent CuO$_{2}$ planes.
Such spin correlations along the c-axis direction would further increase
the size of the correlated clusters shown in Fig. 9.  
For example, if we allow correlations
when any overlap exists among areas of the islands projected to the 
adjacent island on neighbouring planes, the effective 
number of islands contributing to the ``connectivity'' would increase
at least by 3 times compared to the case without interplane correlations.
This is because all the islands on a given CuO$_{2}$ 
plane plus those on the upper and lower planes would participate in the 
correlations.  

In Fig. 10, 
we illustrate this by showing randomly positioned
islands of 30 \% integrated area fraction in three different 
planes (a)-(c).  When we overlap these planes, 
almost all the islands achieve
percolation, as shown in Fig. 10(d).  
Moreover, there could be further contribution from islands
on the second neighbour planes which would enhance connectivity within 
the first neighbour planes.  In this way, we expect a strong tendency
towards percolation of spin correlations among randomly positioned magnetic
islands.

\section{Superconducting Properties}

TF-$\mu$SR measurements allow us to derive the magnetic field penetration 
depth $\lambda$ from the superconducting contribution $\sigma$
of the muon spin relaxation rate as
$\sigma \propto \lambda^{-2} \propto n_{s}/m^{*}$ (superconducting
carrier density / effective mass).  This rate $\sigma$ reflects
the inhomogeneity of the magnetic field in the flux vortex structure,
which varies over a length scale of $\sim$ 1000-3000 \AA.
Any heterogeneity in the superconducting
properties at length scales shorter than this will be averaged out.  
Thus, 
$\mu$SR probes superconductivity with a $\sim$ 100 times coarser
spatial resolution than that with which it probes static magnetism.
In HTSC systems without any static magnetic order, the line shapes in 
TF-$\mu$SR spectra can be used to study more detailed
spatial features, such as the size of the vortex core
region [48].  In the present case, with co-existing 
superconductivity and static magnetic order, however, 
detailed line-shape analyses  
become very difficult.

We have performed TF-$\mu$SR measurements by rotating the incident muon
polarization to be perpendicular to the beam $\vec{P}_{\mu} \perp \hat{z}$
and applying an external field parallel to $\hat{z}$ with the 
c-axis of the specimen mounted either 
parallel to $\hat{z}$ [configuration CT-I]
or perpendicular to $\hat{z}$ [configuration CT-II],
as illustrated in Fig. 11.  The muon spin polarization function $G_{x}(t)$
is measured using two sets of counters placed up (U) and down (D) of the
incident beam.

As shown in
Figs. 12(a) and (b), we observe a larger relaxation in CT-I, the geometry
reflecting the in-plane penetration depth $\lambda_{ab}$, than in CT-II.
We have analyzed the TF-$\mu$SR results by assuming the existence of a static
internal field in a volume fraction $V_{\mu}$ of muon sites,
having the distribution in magnitude and 
direction estimated from the ZF-$\mu$SR measurements.
We have then added vectorially the external field, with a Gaussian
broadening due to superconductivity.  For the 
``zero/low-field muon sites'' with volume fraction (1-$V_{\mu}$),
we assume that the muon sees only the nuclear dipolar fields and the 
Gaussian-broadened external field.
We fit the observed data with this model, and derive the
Gaussian half-width at half maximum of the superconducting
contribution equal to $\sigma/\gamma_{\mu}$.
Figure 12(a) shows the calculated contributions to the relaxation from 
superconductivity and static magnetism, separately.

The temperature dependence of $\sigma$, observed in the CT-I configuration, 
is shown in Fig. 13 for LCO:4.11 and LSCO:0.12.  The relaxation rate shows a
gradual increase with decreasing temperature below $T_{c}$.  
To compare the values of 
$\sigma(T\rightarrow 0)$ with those observed in ceramic specimens of
other HTSC systems [49-52], we multiply by 1/1.4 to account
for the effect of the anisotropic penetration depth [53]. We then
add the corresponding points to the plot of $\sigma(T\rightarrow 0)$
versus $T_{c}$ in Fig. 14.  The points for the present systems lie
on the same trajectory as other LSCO systems. 
We have also included a point for LESCO (Eu0.1, Sr0.15; ceramic), which 
exhibits static magnetic order with $V_{\mu} \sim 50$\% [40,54,55]. 
This point again falls on the trajectory.  The correlation in 
Fig. 14 suggests that $\sigma \propto n_{s}/m^{*}$ at $T\rightarrow 0$
is a determining factor for $T_{c}$ in HTSC systems with 
static stripe freezing (present systems and LESCO) in a way similar to
the case for underdoped [49,50], Zn-doped [52] and overdoped [51,56] 
HTSC systems.

\section{Discussion and Conclusions}

\subsection{Interplay between magnetic and superconducting volumes}

Comparing the
results of $\sigma$ to those observed in ceramic LSCO systems,
assuming that $m^{*}$ is independent of doping, 
the results for LCO:4.11 and LSCO:0.12 correspond to volume average
hole densities of $\sim$ 0.14 +/- 0.02 and 0.10 +/- 0.02 
holes per Cu, respectively.  In view of the large systematic errors, however,
we can not use these results to distinguish whether the volume without
static magnetism (1-$V_{Cu}$) alone or the total volume
carries superconductivity.  
We can rule out, however,
a case where superconducting carriers exist only in the magnetic volume
$V_{Cu}$, since the local hole-concentration for this model would  need to
be unrealistically large (more than 0.3 holes per Cu) 
to account for the observed value of volume integrated $n_{s}/m^{*}$, 
which is comparable to those in LSCO systems in the optimum-doping region. 
So far, a clear indication of 
mutually exclusive regions with static magnetism and
superconductivity has been obtained only in $\mu$SR measurements in 
(La,Eu,Sr)$_{2}$CuO$_{4}$ (LESCO) [40,54,55], which demonstrate that 
the superfluid density $n_{s}/m^{*}$ scales as (1-$V_{\mu}$).  

In previous work on 
La$_{1.45}$Nd$_{0.4}$Sr$_{0.15}$CuO$_{4}$ 
and La$_{1.45}$Nd$_{0.4}$Sr$_{0.2}$CuO$_{4}$ [38], which are superconducting 
below $T_{c} \sim 7$ K and 12 K, respectively, 
$V_{\mu} \sim 100$ \% was found.
This result could be interpreted as evidence of spatial overlap
of regions supporting superconductivity and regions having static 
spin order.  The relationship in Fig. 8(a), however, indicates that
$V_{Cu}$ for this system can be smaller than 100 \%.  This leaves open
the possibility that superconductivity with reduced $T_{c}$
survives in a small remaining volume fraction (1-$V_{Cu}$) $\sim$ 20 \%,
while superconducting regions and regions with static magnetism 
are mutually exclusive.

The situation with magnetic islands, having length scales 
comparable to the in-plane superconducting
coherence length, resembles the ``swiss cheese model'' [52] for Zn-doped
HTSC systems, where a non-superconducting region of comparable size
is created around each Zn.  This model was proposed based on the $\mu$SR
results for the reduction of $n_{s}/m^{*}$ as a function of 
Zn concentration [52],
and was recently confirmed by direct measurements of scanning tunnelling
microscopy (STM) [57].  The $\mu$SR results for overdoped HTSC systems
can also be explained if one assumes spontaneous formation of
hole-rich non-superconducting islands embedded into a sea of hole-poor
superfluid [55, 58-60].  The correlation
between $T_{c}$ and $n_{s}/m^{*}$ survives in all these systems, 
as shown in Fig. 14.
We note that similar microscopic heterogeneity in the 
superconducting state has also been found in 
recent STM measurements on underdoped Bi2212 systems [61].

The $T_{c}$ vs. $n_{s}/m^{*}$
correlations are robust against various perturbations
in HTSC materials, such as those caused by 
Zn-impurities, overdoped fermion carriers, and  the formation of static
magnetic islands, seen in the present work.  This is 
analogous to the robustness of correlations between the  
superfluid transition temperature and the 2-dimensional superfluid density 
in thin films of $^{4}$He and
$^{3}$He/$^{4}$He adsorbed in porous and non-porous media [62-66].
Based on these observations, one of us [55,59,60] pointed out  
a possible relevance of  
these heterogeneous electronic/magnetic features in HTSC systems 
to a ``microscopic phase separation'', such as the one seen in 
superfluid $^{3}$He/$^{4}$He films adsorbed in porous media / fine powders
[66,67].

In the present work on LCO:4.11, as well as a previous $\mu$SR
study on LCO:4.02 and LCO4.04 [43], the
onset of superconductivity occurs around the temperature 
below which the volume fraction of static magnetism 
increases with decreasing temperature.  This apparent 
coincidence of $T_{c}$ and $T_{N}$ could be expected if
a microscopic rearrangement of charge distribution occurs at $T_{c} \sim T_{N}$
to separate the electron systems into hole-rich regions
which support superconductivity and hole-poor ($x \sim 1/8$
for LCO:4.11 and $x \sim 0$ for LCO:4.02 and LCO:4.04) regions
which support static magnetism.  

Recently, Kivelson {\it et al.\/} [68] have
proposed a model with a particular type of phase separation to
explain the sharp rise of $\nu$ and gradual increase of 
$V_{Cu}$ below $T_{N}$ seen in the present study of LCO:4.11.
Their model also predicts a trade-off of
superconducting and magnetic volumes below $T_{c} \sim T_{N}$.
The gradual change of both $\sigma(T)$ and $V_{\mu}(T)$ below
$T_{c} \sim T_{N}$ in LCO:4.11 (Figs. 3(a) and 13),
however, seems inconsistent with such a trade-off.  

The superconducting $T_{c}$ and the magnetic $T_{N}$ appear at different
temperatures in $\mu$SR measurements of LSCO:0.12 as well 
as in various LNSCO and LESCO
systems, with  LSCO:0.12 having $T_{c} > T_{N}$
and the other materials having $T_{N} > T_{c}$.   It should be 
noted that neutron measurements give 
$T_{c} \simeq T_{N}$ in LSCO:0.12, while $T_{N}$ determined by $\mu$SR
is lower, presumably due to difference in time windows between
neutron and muon measurements.  In these systems the frequency
$\nu$ increases with decreasing temperature gradually below $T_{N}$,
as shown in Fig. 3(b), and in refs. [38-42].
Thus, the abrupt development of the magnetic order parameter is not a common
feature of all the HTSC systems.

\subsection{Bragg peak intensity in LCO:4.11 and AF-LCO}

We can estimate the Bragg peak intensity
$I_{B:4.11}$ expected in neutron scattering measurements on LCO:4.11
from the $\mu$SR results.
$I_{B:4.11}$
should be reduced from the value $I_{B:AF}$ in AF-LCO 
by a factor 0.5 for the SDW amplitude ($sin^{2}$) 
times the ratio of the maximum static moments (0.36/0.55)$^{2}$
times the volume fraction of magnetic Cu atoms $V_{Cu} = 0.27$.  
Thus, we predict as a rough estimate $I_{B:4.11} = 0.06 I_{B:AF}$.  
This value agrees well
with the observed neutron results $I_{B:4.11} = 0.07 I_{B:AF}$.
In Reference [16] it is assumed that three out of four copper 
ions are magnetic and that they are distributed uniformly 
throughout the entire volume, so a smaller average moment 
per Cu$^{2+}$ of 0.15 $\mu_{B}$ is inferred.

We note, however, that estimates of the ordered moment size from
neutron studies and from $\mu$SR often disagree with one another.
For example, the N\'eel temperature in antiferromagnetic 
La$_{2}$CuO$_{4+\delta}$
varies from $T_{N} \sim 300$ K to $T_{N} \leq 100$ K with 
a small variation in $\delta$.  In neutron 
scattering studies of these AF-LCO systems, the Bragg peak intensity
$I_{B}$ shows a reduction, by more than a factor of 10, 
with decreasing $T_{N}$ [28].
In contrast, the frequency of ZF-$\mu$SR spectra shows only a 20 \%\
change in the local frozen moment for the same reduction of $T_{N}$.  
These $\mu$SR
and neutron results on AF-LCO are compared in ref. [28].
Since the $\mu$SR frequency is directly proportional to the magnitude of 
neighbouring static Cu moments, the reduction of $I_{B}$ in neutron
studies cannot be ascribed to a change of individual Cu moment size.
The likely reason for the strong reduction of $I_{B}$ is a 
progressive trade-off between the true long range ordered component and 
short range
spin glass fluctuations which recent work [69] has shown 
involves fluctuations into the diagonal
stripe spin glass phase with hole concentration of $\sim 0.02$.

\subsection{Neutron results in high magnetic fields}

Recently Khaykovich {\it et al.\/} [48] 
have made measurements on LCO:4.11
in high magnetic fields $H_{ext}$ applied parallel to
the c-axis.  By increasing the external field
from $H_{ext} = 0$
to $H_{ext} = 9$ T, 
they have found a factor $\sim 2$ increase of 
the intensity $I_{B}$ of the magnetic Bragg peak
at $T\rightarrow 0$ for a crystal similar to the one studied here.  
The superconducting $T_{c}$ is suppressed
by $H_{ext}$, while $T_{N}$ remains nearly unchanged up to 
$H_{ext} = 9$ T.  

Application of $H_{ext}$ perpendicular to the CuO$_{2}$ plane
creates vortices.  The vortex core, with radius comparable to 
the in-plane coherence length $\xi_{ab}\sim 30$ \AA\, becomes
normal, as illustrated in Fig. 15.  These normal cores could
have static magnetic order similar to that in the 
surrounding magnetic islands.   
 
The fraction of the area contained in vortex cores may be estimated 
from the upper 
critical field, since that is the field at which the cores fill 
the entire area.  
Using $H_{c2} \geq 40$ T, found for optimally doped LSCO [70],  we expect that
only $\sim 1/4$ or less of the area that is superconducting at $H_{ext} = 0$
would be turned into a normal core at $H_{ext} = 9$ T.  
However, this is sufficient to explain the factor $\sim2$ 
increase in the Bragg peak intensity if the zero-field 
sample has ordered moments in only $\sim27$\%\ of its volume and if 
the magnetism in the cores is coherent with the long range 
order present at zero field.  We are currently undertaking $\mu$SR 
studies of LCO:4.11
in high applied magnetic fields, which will be reported in future publications.

In nearly optimally doped LSCO, Lake {\it et al.\/} [71] have recently
found that low-energy neutron scattering intensity, 
below the energy transfer corresponding to the superconducting energy gap,
increases at temperatures well below $T_{c}$, when a high external
field $H_{ext}$ is applied along the c-axis.  This phenomenon may
also be related to the magnetic response from the flux-core regions,
since the increase of the intensity sets in at the flux-depinning
temperature.
A rather surprising feature of this study is the sharpness of the
observed low-energy fluctuations in momentum space, suggesting 
long-range spin correlations
among widely separated vortex core regions.
We point out here that, like the static order, this requires long 
range coupling between vortex cores.

\subsection{Conclusions}

In summary, we have performed ZF-$\mu$SR measurements using 
LCO:4.11 and LSCO:0.12 single crystals, and have found that static
incommensurate SDW spin freezing, with the maximum ordered
Cu moment size of 0.36 $\mu_{B}$, develops only in a partial site fraction.
We assume $V_{\mu}\sim$ 40\%\ for LCO:4.11 and $\sim$ 18\%\ for 
LSCO:0.12, and specific modeling suggests that the 
corresponding ordered Cu fractions may be somewhat smaller.  
Comparison of observed results with computer
simulation suggests the formation of static magnetic islands 
on the CuO$_{2}$ planes having size $R = 15\sim 30$ \AA,
comparable to the in-plane
superconducting coherence length $\xi_{ab}$.  Order between these 
islands nevertheless propagates over long distances to yield a 
correlation length in excess of 600 \AA.

We stress that LCO:4.11, which is a stoichiometric single 
crystal, has the least
built-in randomness among the various HTSC systems that exhibit 
static incommensurate magnetic correlations.  Yet the present
work demonstrates that the ground state is a mixture of
magnetically ordered and at best weakly ordered regions.  
Near phase boundaries
or quantum critical points in strongly
correlated electron systems, such microscopic heterogeneity could 
result from competition of the different 
order parameters involved.  Related phenomena have been observed,
for example, in the formation of 
stripe magnetic correlations in manganites [72] and the development of
stripe domains in 
LSCO with $x \sim 0.01$ to 0.02 [73].
These observations encourage further theoretical studies
of electronic states involving spontaneous
formation of heterogenous regions in competing order parameter systems.

Our TF-$\mu$SR measurements in these systems 
demonstrate that $n_{s}/m^{*}$ at $T\rightarrow 0$ 
exhibits correlations with $T_{c}$ not only in underdoped,
Zn-doped, and overdoped HTSC systems but also in the present
systems with static SDW spin freezing.  Recent measurements of 
the penetration depth in (BEDT-TTF)$_{2}$Cu(NCS)$_{2}$ in applied
pressure [74], and in A$_{3}$C$_{60}$ systems [75,76], suggest that
these correlations are followed also by organic and fullerine
superconductors. 

We have proposed a model for explaining the rather long-range
spin correlations in these systems, resorting to ``connectivity''
of neighbouring magnetic islands on the same and adjacent
CuO$_{2}$ planes.  We have also suggested that static magnetism in the 
vortex core region can provide  
explanations for the 
field dependence of the neutron scattering results in LCO:4.11
and in optimally doped LSCO systems.

The present work demonstrates a very important feature of $\mu$SR,
i.e., the capability to determine the volume fraction of magnetically
ordered regions.  Recently, a similar case was noticed in the
study of URu$_{2}$Si$_{2}$, where static magnetism below $T_{N} \sim 17$ K
has been identified to exist in a partial volume fraction by combination
of neutron [77] and NMR [78] 
studies in applied pressure.  The
first evidence for this feature was indeed provided by a $\mu$SR 
measurement [79] in ambient pressure.  In the $\mu$SR results 
for URu$_{2}$Si$_{2}$, Luke {\it et al.} found a very sharp onset 
of the precession frequency $\nu(T)$ below $T_{N}$, 
temperature dependence of the precessing amplitude $V_{\mu}(T)$, 
and a good agreement
between the temperature dependence $I_{B}(T)$ of the neutron Bragg
peak and $V_{\mu}\nu^{2}$ from $\mu$SR.  These features are common to the
present results in LCO:4.11.  We also note that 
ZF-$\mu$SR results in CeCu$_{2.2}$Si$_{2}$ [80,81] 
indicate a possible trade-off between magnetic and superconducting
volume fractions. These results suggest possible involvement of 
microscopic phase separation in both HTSC systems and heavy fermion systems
[55,68,82].  Further detailed studies of $\mu$SR in combination with 
neutron scattering will be very helpful in obtaining an
overall understanding of the interplay between magnetism and superconductivity
in HTSC, heavy-fermion, and other strongly correlated electron systems.  

\section{Acknowledgement}

We wish to acknowledge helpful discussions with B. Nachumi, G. Shirane, 
J.M. Tranquada and S. Uchida.
The work at Columbia was supported by
NSF (DMR-98-02000 and DMR-01-02752) and US-Israel Binational
Science Foundation.  Research at McMaster is supported by NSERC and 
the Canadian Institute for Advanced Research.
The work at MIT has
been supported by the MRSEC Program of the National Science
foundation under Award No. DMR 9808941 and
 by NSF under Awards No.
DMR 0071256 and DMR 99-71264. Work at the University of Toronto is
part of the Canadian Institute for Advanced Research and is
supported by the Natural Science and Engineering Research Council
of Canada.

\newpage

\begin{figure}

\begin{center}

\mbox{\psfig{figure=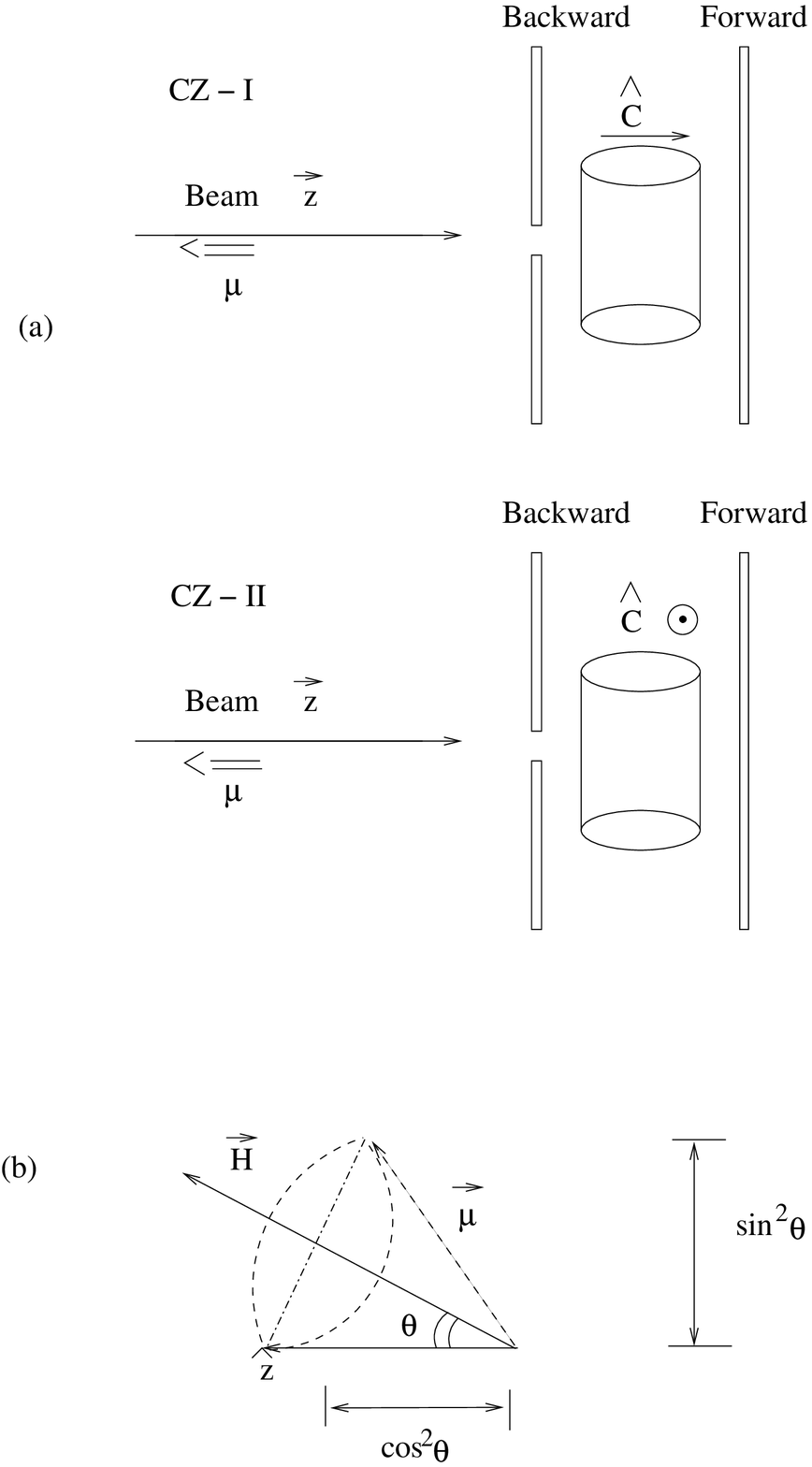,width=3in}}

\vskip 0.2 truein

\label{Figure 1.} 

\caption{
(a) Schematic view of the experimental
configurations in zero field $\mu$SR (ZF-$\mu$SR) employed in the 
present study.  Measurements were performed
with the initial muon spin polarization parallel (CZ-I)
and perpendicular (CZ-II) to 
the $\hat c$ - axis, by rotating the crystal orientation by 
90 degrees.  The upper [lower] figure shows the CZ-I [CZ-II] 
configuration. (b) Illustration of the
oscillating part of the muon polarization signal, proportional to
$sin^{2}(\theta) $, with $\theta $ being the angle between the local
magnetic field and the initial muon spin direction.  The  
$cos^{2}(\theta)$ component contributes to the amplitude of the
non-oscillating signal.}
\end{center}
\end{figure}
\newpage

\begin{figure}

\begin{center}

\mbox{\psfig{figure=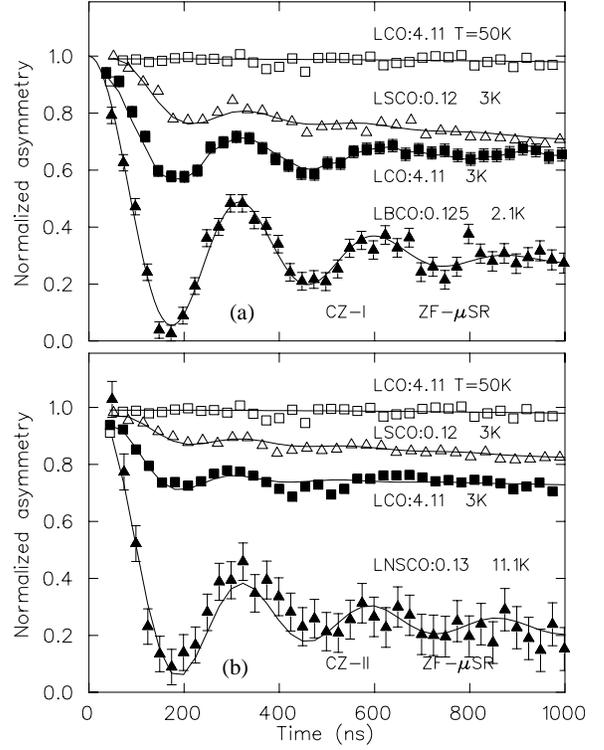,width=3in}}

\vskip 0.2 truein

\label{Figure 2.} 

\caption{
Time spectra of the muon spin polarization 
observed by ZF-$\mu$SR with (a) the CZ-I configuration, 
and (b) the CZ-II configuration for
La$_{2}$CuO$_{4.11}$ (LCO:4.11)
and La$_{1.88}$Sr$_{0.12}$CuO$_{4}$ (LSCO:0.12).
Also included are results for 
ceramic specimens of La$_{1.875}$Ba$_{0.125}$CuO$_{4}$ (LBCO:0.125)
from [36] 
and La$_{1.47}$Nd$_{0.4}$Sr$_{0.13}$CuO$_{4}$ (LNSCO:0.13) from [38]}

\end{center}

\end{figure}

\newpage

\begin{figure}

\begin{center}

\mbox{\psfig{figure=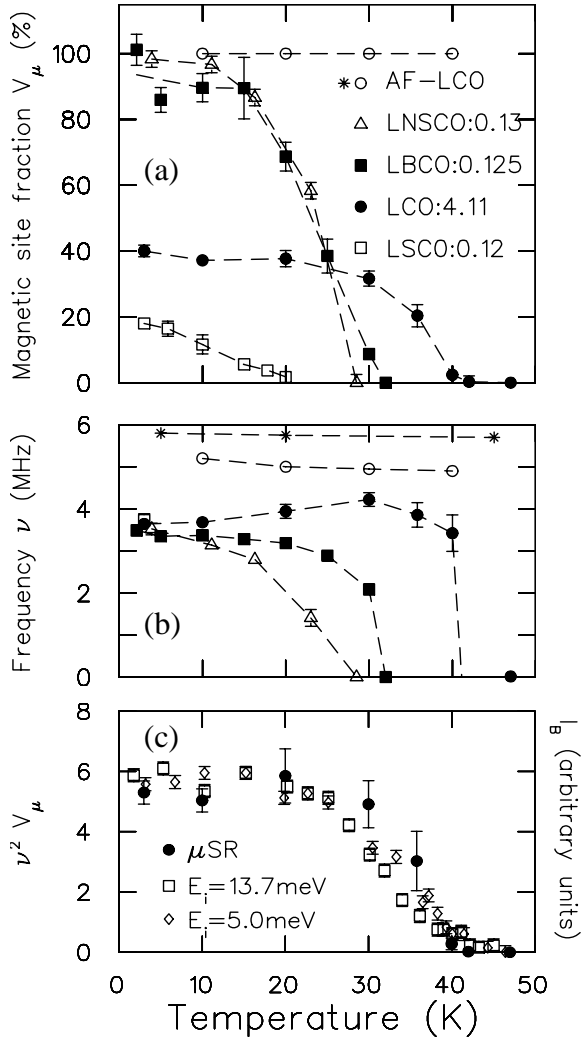,width=3in}}

\vskip 0.2 truein

\label{Figure 3.} 

\caption
{
(a) Volume fraction $V_{\mu}$ of muon sites
with a static magnetic field larger than $\sim 30$ G
and (b) frequency of the precessing signal in La$_{2}$CuO$_{4.11}$ (LCO:4.11)
and La$_{1.88}$Sr$_{0.12}$CuO$_{4}$ (LSCO:0.12) (from the present study),
compared with the results in 
La$_{1.875}$Ba$_{0.125}$CuO$_{4}$ (LBCO:0.125) [36],
La$_{1.47}$Nd$_{0.4}$Sr$_{0.13}$CuO$_{4}$ (LNSCO:0.13) [38], and
antiferromagnetic La$_{2}$CuO$_{4+\delta}$ (AF-LCO) [27,28].
The broken lines are guides to the eye.
(c) Comparison of the temperature dependence of the Bragg peak intensity 
$I_{B}$ of neutron scattering measurements in La$_{2}$CuO$_{4.11}$ (LCO:4.11)
[16] with those expected from the $\mu$SR results (present study)
as $I_{B} \propto V_{\mu} \times \nu^{2}$.  $\mu$SR and neutron 
results are scaled using the values near $T\rightarrow 0$.  
Comparison of the absolute values is discussed in Section V-B.
}

\end{center}

\end{figure}

\newpage

\begin{figure}

\begin{center}

\mbox{\psfig{figure=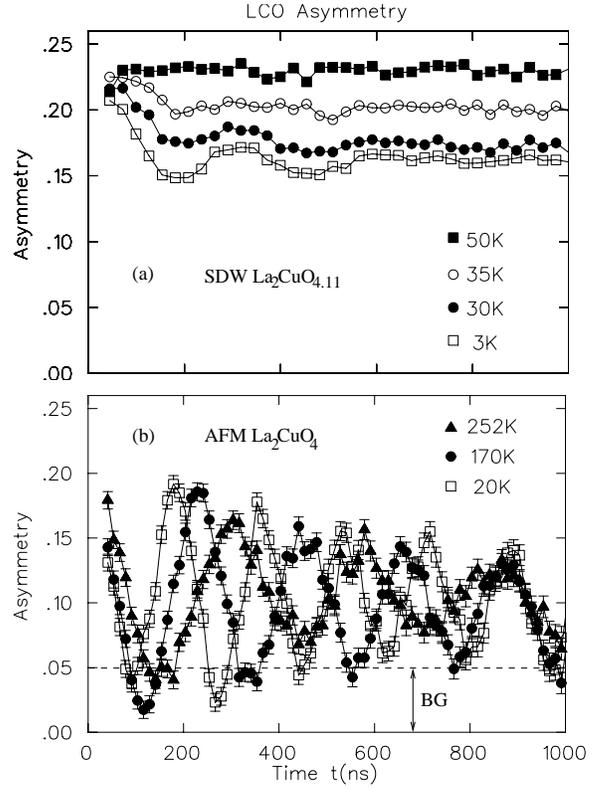,width=3in}}

\vskip 0.2 truein

\label{Figure 4.} 

\caption
{
Time spectra of ZF-$\mu$SR measurements in (a) LCO:4.11
and (b) AF-LCO, at various temperatures.
The amplitude of the oscillating and relaxing signal varies
in (a) below $T_{N}$ without much change in the 
frequency, while in (b) the amplitude does not
depend on $T$ and the frequency increases with decreasing
$T$.  The latter behavior is observed in ZF-$\mu$SR
of many conventional magnetic systems.  The spectrum (a)
was obtained by using low-background $\mu$SR spectrometer,
which ensures that the background signal level is below 0.01 in 
Asymmetry.  The spectrum (b) was obtained in 1987 by using
older apparatus.  The broken line (BG) indicates an expected level
of background signal in (b).
}

\end{center}

\end{figure}

\newpage

\begin{figure}

\begin{center}

\mbox{\psfig{figure=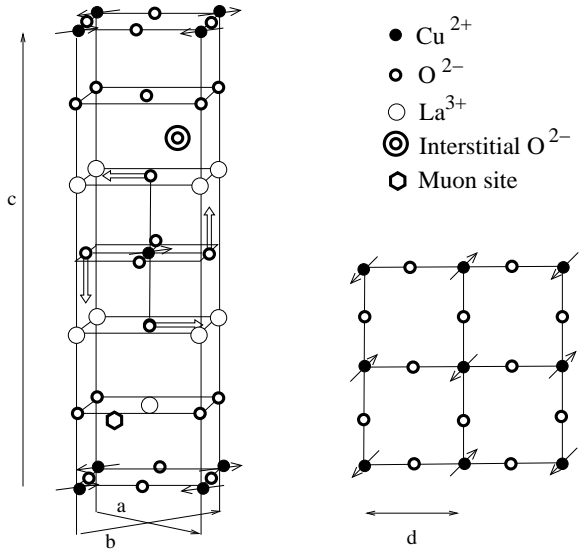,width=3in}}

\vskip 0.2 truein

\label{Figure 5.} 

\caption
{
Schematic view of the crystal structure of 
La$_{2}$CuO$_{4}$, shown with the positions of intercalated interstitial
oxygen and the muon site associated with an apical oxygen. 
We assume the location of muon site near apical oxygen
to be (0,0.751,2.257) \AA\  within the unit cell.
The arrows attached to O$^{2-}$ in the center of the figure show
the direction of tilting (about 5$^{o}$) of the CuO$_{6}$ octahedra.  
}

\end{center}

\end{figure}

\newpage

\begin{figure}

\begin{center}

\mbox{\psfig{figure=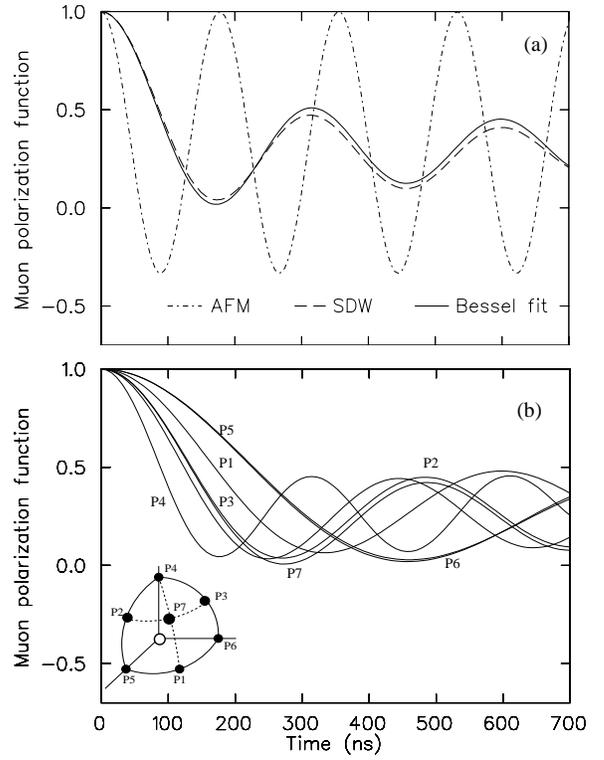,width=3in}}

\vskip 0.2 truein

\label{Figure 6.} 

\caption
{
Computer simulation results of the
muon spin polarization functions expected for La$_{2}$CuO$_{4}$
assuming that all the Cu moments involved in the static magnetic order.
An angular average is taken to represent the results for ceramic
specimens. (a) Results for the muon site near the apical
oxygen (see Fig. 5). 
The dotted line corresponds to the
case with antiferromagnetic (AFM) spin correlations with 
the static Cu moment of 0.55 $\mu_{B}$, 
while the dashed line corresponds to the case with spin density wave (SDW)
amplitude modulation with maximum static
Cu moment of 0.35 $\mu_{B}$ and modulation 
wavevector $k = 0.12*2\pi /d$ (d=3.779 \AA) superimposed on the 
antiferromagnetic correlations.
The solid line is the result of a fit
to a Bessel function. (b) Results for several
hypothetical muon sites near the intercalated interstitial oxygen,
illustrated in the inset, obtained for the above-mentioned
SDW spin correlations.
}

\end{center}

\end{figure}

\newpage

\begin{figure}

\begin{center}

\mbox{\psfig{figure=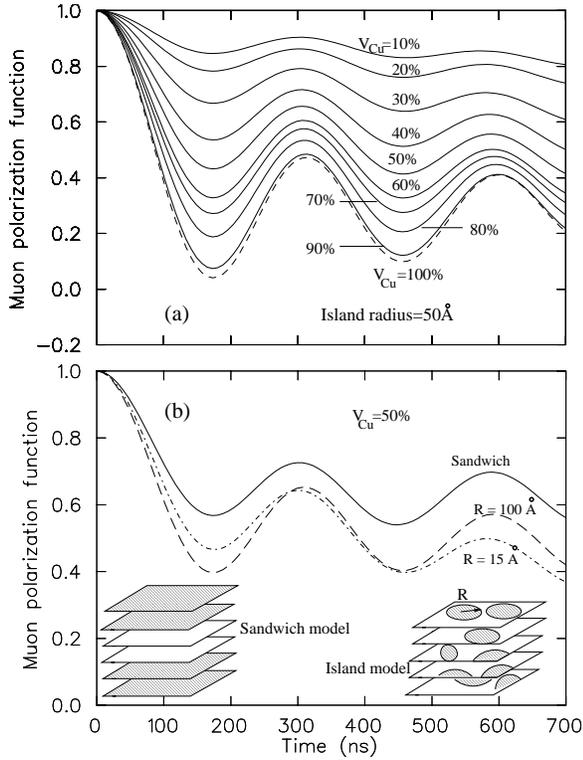,width=3in}}

\vskip 0.2 truein

\label{Figure 7.} 

\caption
{
Computer simulation results of the
expected muon spin polarization functions for La$_{2}$CuO$_{4}$,
with only a part of the Cu moments involved in the static magnetic order,
obtained for the muon site near the apical oxygen 
after angular average to simulate the results for ceramic specimens.
We assume a
SDW modulation with a maximum static Cu moment of 0.35 $\mu_{B}$ and
modulation vector $k = 0.12\times 2\pi/d$.
(a) Magnetic island
model where the Cu moments in a
volume fraction $V_{Cu}$ order with
the SDW amplitude modulation, forming islands of radius
$R = 50$ \AA. (b) Comparison of the results for the magnetic island model
with $R = $15 and 100 \AA\ with those for the 
``sandwich model'' where two CuO$_{2}$ planes adjacent to the 
intercalated oxygen layer do not have any static Cu moments while
all the 
Cu moments on the other two planes are ordered.  All three curves shown in (b) correspond to $V_{Cu}$=50{\%}. The
insets illustrate these two models, 
with the shaded regions containing SDW modulated static 
Cu moments and the blank regions without static Cu moments.
}

\end{center}

\end{figure}

\newpage

\begin{figure}

\begin{center}

\mbox{\psfig{figure=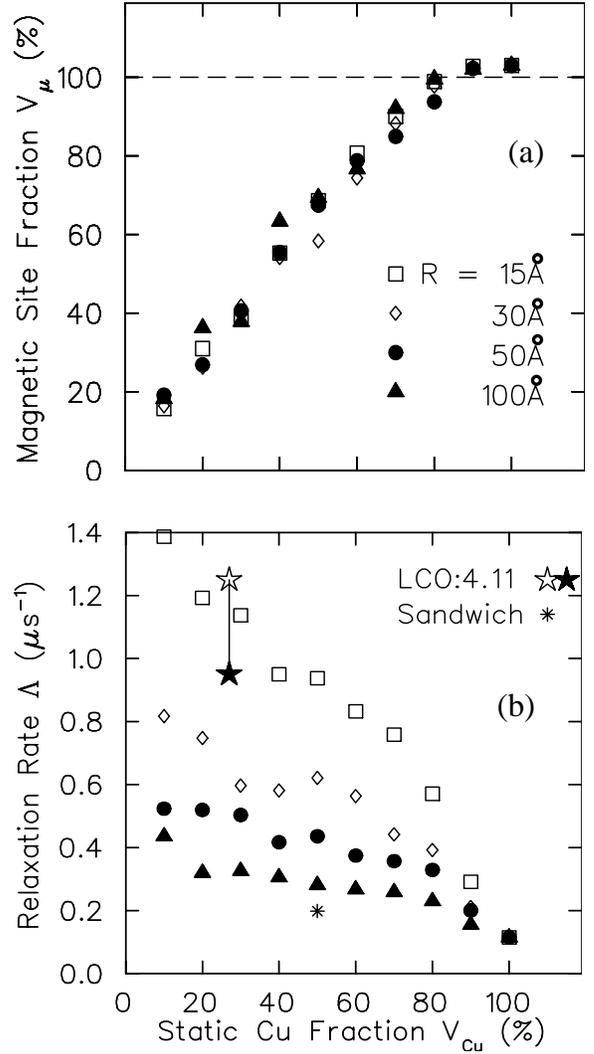,width=3in}}

\vskip 0.2 truein

\label{Figure 8.} 

\caption
{
(a) Computer simulation results for the volume fraction $V_{\mu}$ 
of muon sites
with a static magnetic field larger than $\sim 30$ G as a function of
the volume fraction of the sample containing static Cu moments, $V_{Cu}$.
(b) The relaxation rate $\Lambda$ of the Bessel function 
oscillation, obtained by fitting the simulation results with Eq. 3, 
plotted as a function of $V_{Cu}$.  Comparison 
with the experimental results for LCO:4.11 (Raw data
$\Lambda_{4.11}$ shown by an open star symbol and corrected data
$\Lambda_{4.11}^{c}$ by a closed star symbol) allows 
estimation of the size of magnetic islands.   The asterisk symbol *
shows the relaxation rate $\Lambda$ expected for the sandwich model.
}

\end{center}

\end{figure}

\newpage

\begin{figure}

\begin{center}

\mbox{\psfig{figure=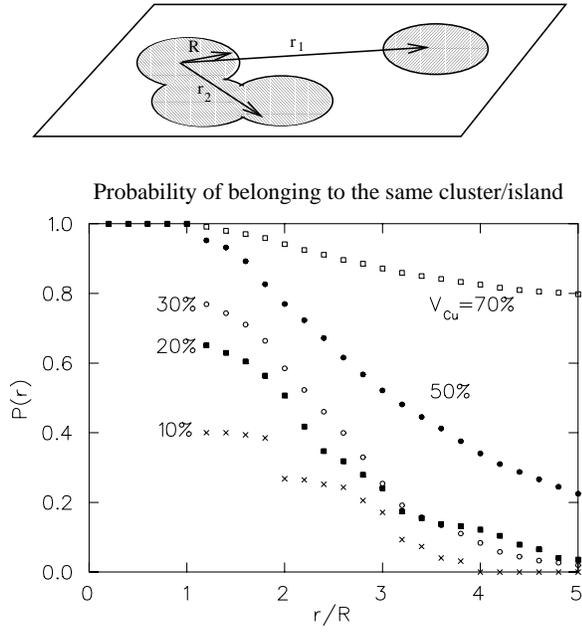,width=3in}}

\vskip 0.2 truein

\label{Figure 9.} 

\caption
{
Probability $P(r)$ that Cu spins belong to a certain cluster versus the
distance from the center of an of radius R in the cluster. 
We assume 
that all the spins in a given island are correlated,
and the correlation extends if there is any small overlap
between adjacent islands on a given CuO$_{2}$ plane,
as illustrated in the upper figure.  For example, a Cu spin 
at a distance $r_{2}$ contributes ``1.0'' for the probability,
while that at $r_{1}$ gives ``0.0'' contribution.  We average
over all the possible cluster/island configurations to calculate
$P(r)$.  
The results demonstrate a long length scale of spin correlations 
for the ``magnetic island model'', due to formation of 
large clustered islands.
}

\end{center}

\end{figure}

\newpage

\begin{figure}

\begin{center}

\mbox{\psfig{figure=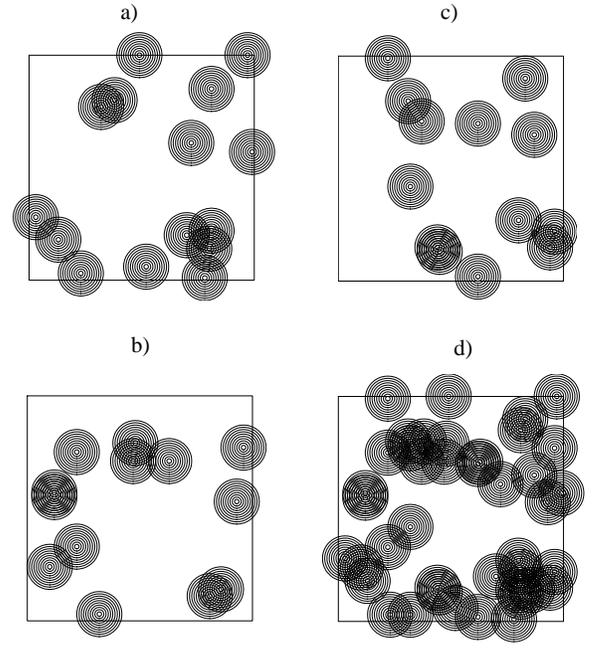,width=3in}}

\vskip 0.2 truein

\label{Figure 10.} 

\caption
{
Illustration of percolating cluster islands.  (a)-(c)
show planes with random locations of magnetic islands
having integrated area fraction of 30 \%. (d) shows the overlap
of (a)-(c).  (d) demonstrates that 
most of the islands belong to the percolating cluster if we allow correlations of spins in ``overlapping islands''
on all the three planes.}

\end{center}

\end{figure}

\newpage

\begin{figure}

\begin{center}

\mbox{\psfig{figure=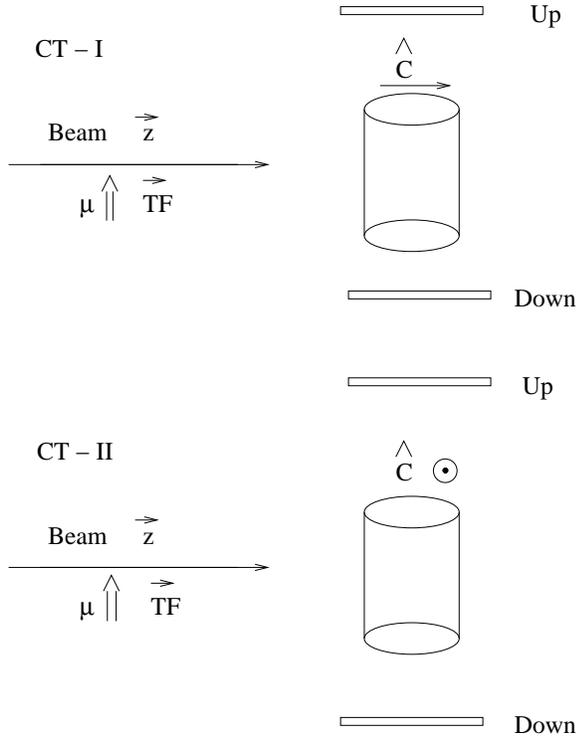,width=3in}}

\vskip 0.2 truein

\label{Figure 11.} 

\caption
{
Schematic view of 
experimental configurations of Transverse Field $\mu$SR 
(TF-$\mu$SR) measurements employed in the present work with the 
external field applied 
parallel (CT-I) and perpendicular (CT-II) to the $\hat c$-axis
of the single crystal.
}

\end{center}

\end{figure}

\newpage

\begin{figure}

\begin{center}

\mbox{\psfig{figure=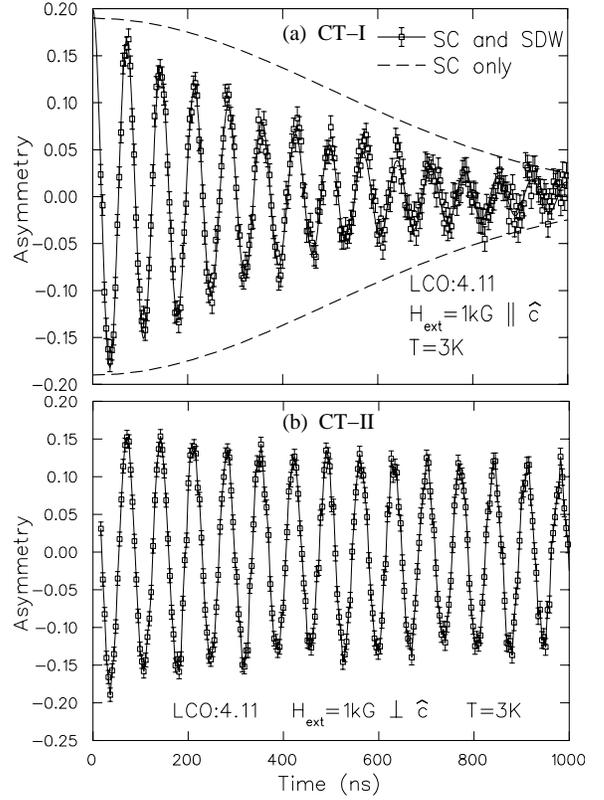,width=3in}}

\vskip 0.2 truein

\label{Figure 12.} 

\caption
{
Time spectra of the muon 
spin polarization observed in TF-$\mu$SR in LCO:4.11
obtained using (a) CT-I and (b) CT-II configurations. 
The relaxation of the signal is due to a distribution of
magnetic fields inside a superconducting sample (SC) and magnetic fields due
to static Cu moments with the SDW amplitude modulation.
(a) shows faster damping than (b), reflecting the shorter penetration 
depth for the external field applied perpendicular to the CuO$_{2}$
planes.  
The expected effect from superconductivity alone is 
shown by the dotted line in (a).
}

\end{center}

\end{figure}

\newpage

\begin{figure}

\begin{center}

\mbox{\psfig{figure=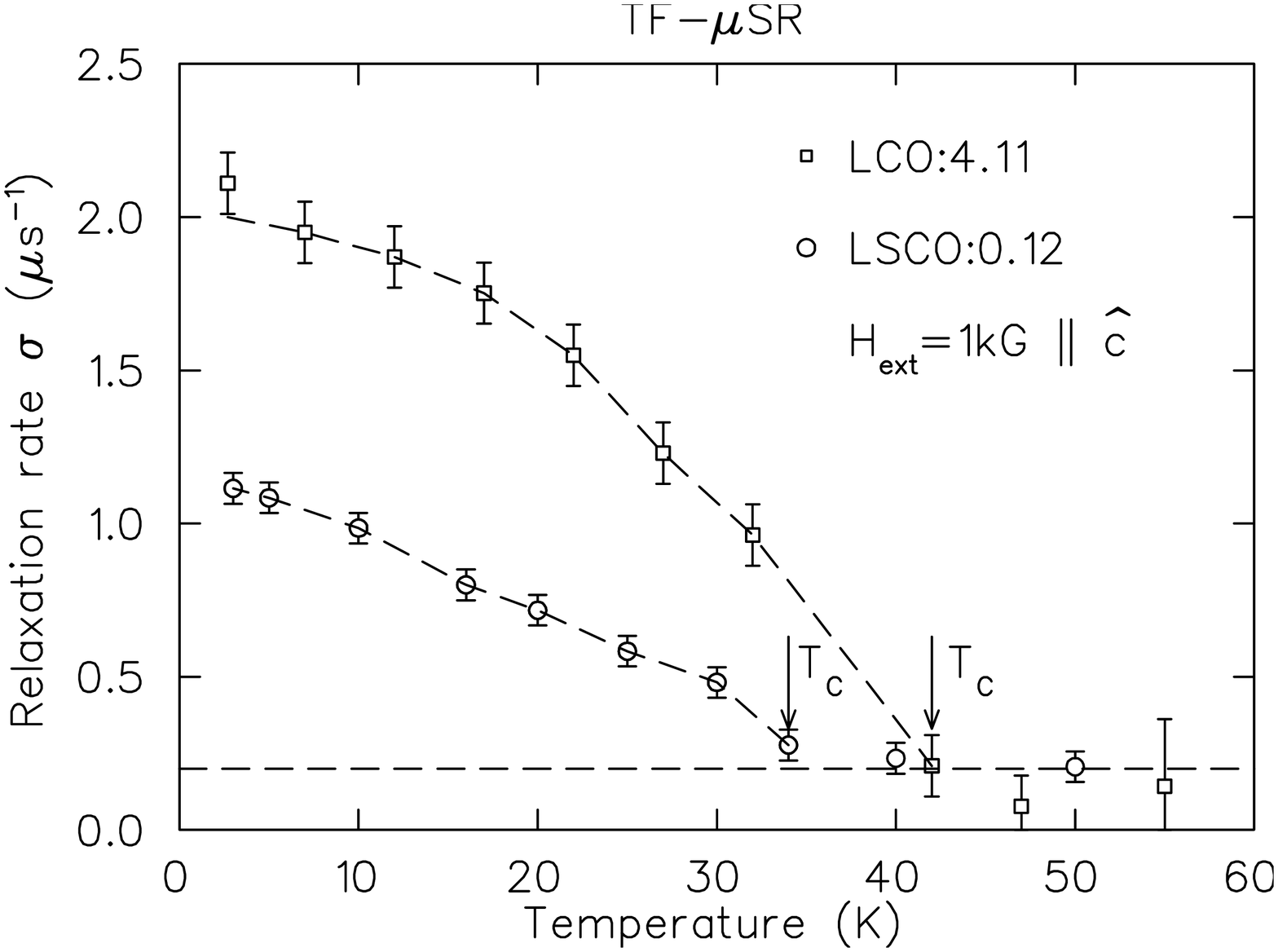,width=3in}}

\vskip 0.2 truein

\label{Figure 13.} 

\caption
{
Temperature dependence of the 
superconducting relaxation rate $\sigma$ due to superconducting flux vortices, 
observed in TF-$\mu$SR measurements (with the CT-I configuration)
in La$_{2}$CuO$_{4.11}$ (LCO:4.11) and La$_{1.88}$Sr$_{0.12}$CuO$_{4}$
(LSCO:0.12). The broken lines are guides to the eye.
}

\end{center}

\end{figure}

\newpage

\begin{figure}

\begin{center}

\mbox{\psfig{figure=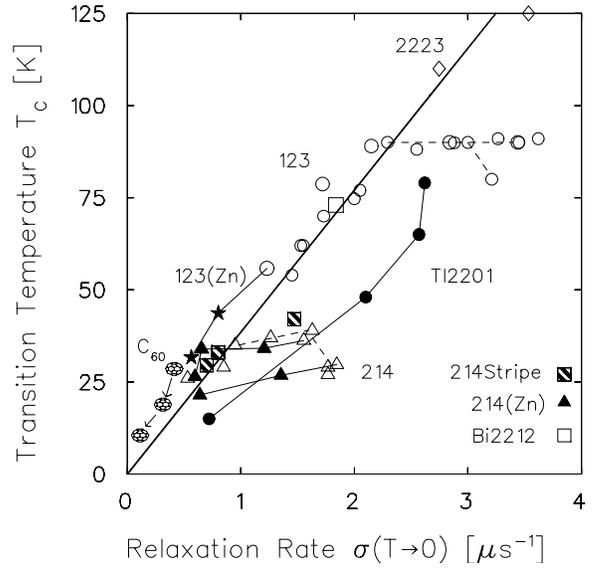,width=3in}}

\vskip 0.2 truein

\label{Figure 14.} 

\caption
{
A plot of the superconducting transition temperature $T_{c}$
versus the 
relaxation rate $\sigma(T\rightarrow 0)$ at low temperatures 
(proportional to the superfluid density
$n_{s}/m^{*}$) for several high-$T_{c}$ superconductors
[49-52]. The results with the
``stripe square'' symbols represent points from
LESCO ([54,55]), 
LSCO:0.12, and LCO:4.11 in the order of increasing $\sigma(T\rightarrow 0)$.
To account for difference between results for ceramic and single-crystal
specimens, the values for $\sigma$ for LCO:4.11 and LSCO:0.12
in the CT-I configuration were multiplied by 1/1.4.
}

\end{center}

\end{figure}

\newpage

\begin{figure}

\begin{center}

\mbox{\psfig{figure=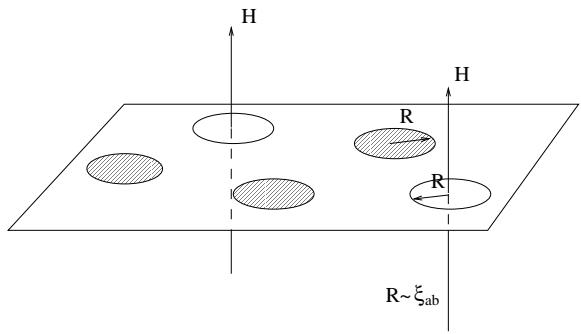,width=3in}}

\vskip 0.2 truein

\label{Figure 15.} 

\caption
{
An illustration of magnetic islands (shaded circles) 
and vortex cores (open circles) on the
CuO$_{2}$ planes of HTSC systems under an external magnetic
field applied along the c axis direction.
When a core is located at the magnetic island, one expects 
an increase in the radius of the non-superconducting region 
due to supercurrent flowing around such a magnetic island.}

\end{center}

\end{figure}


\begin{thebibliography}{99}
\def\eja{E.J.~Ansaldo}
\def\gae{G.~Aeppli}
\def\jda{J.D.~Axe}
\def\msa{M.S.~Alvarez}
\def\iaf{I.~Affleck}
\def\pea{P.E.~Armstrong}
\def\yaj{Y.~Ajiro}
\def\jhb{J.H.~Brewer}
\def\bba{B.~Batlogg}
\def\rjb{R.J.~Birgeneau}
\def\pbo{P.~B\"oni}
\def\ebu{E.~Bucher}
\def\sba{S.~Barth}
\def\cba{C.~Baines}
\def\pba{P.~Barboux}
\def\jwb{J.W.~Brill}
\def\dab{D.A.~Bonn}
\def\tcs{T.~Csiba}
%
\def\wsc{W.S.~Chan}
\def\sch{S.~Chikazumi}
\def\cwc{C.W.~Clawson}
\def\rac{R.A.~Cowley}
\def\kmc{K.M.~Crowe}
\def\jec{J.E.~Crow}
\def\hsc{H.S.~Chen}
\def\mce{M.~Celio}
\def\jfc{J.F.~Carolan}
\def\acd{A.C.D.~Chaklader}
\def\dec{D.E.~Cox}
\def\clc{C.L.~Chien}
\def\mzc{M.Z.~Cieplak}
\def\brc{B.R.~Cyca}
\def\lyc{L.Y.~Chiang}
%
\def\pmc{P.M.~Chaikin}
\def\kdc{K.D.~Carlson}
\def\amd{A.M.~de~Graaf}
\def\mdo{M.~Doyama}
\def\ldo{L.~Dobrzynski}
\def\pdo{P.~Dosanjh}
\def\mrd{M.R.~Davis}
\def\yda{Y.~Dalichaouch}
\def\hdr{H.~Drulis}
\def\fdi{F.~Diederich}
\def\sdo{S.~Donovan}
\def\yen{Y.~Endoh}
\def\vje{V.J.~Emery}
\def\yeno{Y.~Enomoto}
\def\set{S.~Etemad}
\def\eae{E.A.~Early}
\def\rwe{R.W.~Ellis}
%
\def\dgf{D.G.~Fleming}
\def\hfu{H.~Fukushima}
\def\yfu{Y.~Fukai}
\def\zfi{Z.~Fisk}
\def\dpg{D.P.~Goshorn}
\def\fng{F.N.~Gygax}
\def\jgo{J.~Goparakrishnan}
\def\pmg{P.M.~Gehring}
\def\jdg{J.D.~Garrett}
\def\ggr{G.~Gr\"uner}
\def\cyh{C.Y.~Huang}
\def\rsh{R.S.~Hayano}
\def\drh{D.R.~Harshman}
\def\tha{T.~Hatano}
\def\jjh{J.J.~Hauser}
\def\hgu{H.~Guggenheim}
\def\bhi{B.~Hitti}
%
\def\wnh{W.N.~Hardy}
\def\meh{M.E.~Hayden}
\def\yhi{Y.~Hidaka}
\def\hrh{H.R.~Hart}
\def\ths{T.~Hsu}
\def\dgh{D.G.~Hinks}
\def\khw{K.~Hepburn-Wiley}
\def\smh{S.M.~Huang}
\def\kho{K.~Holczer}
\def\yis{Y.~Ishikawa}
\def\jim{J.~Imazato}
\def\mit{M.~Itoh}
\def\mis{M.~Ishikawa}
\def\sis{S.~Isibashi}
\def\vja{V.~Jaccarino}
\def\dcj{D.C.~Johnston}
\def\ajj{A.J.~Jacobson}
\def\jdj{J.D.~Jorgensen}
%
\def\wka{W.~Kang}
\def\mki{M.~Kinoshita}
\def\tko{T.~Komatsubara}
\def\cmk{C.M.~Krowe}
\def\rfk{R.F.~Kiefl}
\def\pki{P.~Kienle}
\def\hko{H.~Koyama}
\def\yku{Y.~Kuno}
\def\rku{R.~Kubo}
\def\sek{S.E.~Kohn}
\def\yki{Y.~Kitaoka}
\def\rka{R.~Kadono}
\def\rke{R.~Keitel}
\def\wjk{W.J.~Kossler}
\def\jrk{J.R.~Kempton}
\def\kkat{K.~Katsumata}
\def\srk{S.R.~Kreitzman}
\def\nka{N.~Kaplan}
\def\aku{A.~Kupla}
\def\dke{D.~Keane}
\def\kka{K.~Kakurai}
\def\rkr{R.~Krahn}
\def\hkoj{H.~Kojima}
\def\amk{A.M.~Kini}
\def\ake{A.~Keren}
\def\rbk{R.B.~Kaner}
\def\kko{K.~Kojima}
\def\gml{G.M.~Luke}
%
\def\wfl{W.F.~Lankford}
\def\jtl{J.T.~Lewandowski}
\def\kwl{K.W.~Ley}
\def\vle{V.Y.~Lee}
\def\lpl{L.P.~Le}
\def\qli{Q.~Li}
\def\bwl{B.W.~Lee}
\def\sol{Sophia~Lin}
\def\hmi{H.~Miyajima}
\def\cfm{C.F.~Majkrzak}
\def\pwm{P.W.~Mitchell}
\def\apm{A.P.~Murani}
\def\arm{A.R.~Moodenbaugh}
\def\hmit{H.~Miyatake}
\def\tmu{T.~Murakami}
\def\pmu{P.~Mulhern}
\def\mbm{M.B.~Maple}
\def\llm{L.L.~Miller}
\def\jtm{J.T.~Markert}
\def\mme{M.~Mekata}
\def\kni{K.~Nishiyama}
\def\kna{K.~Nagamine}
\def\rna{R.~Nakai}
\def\nni{N.~Nishida}
\def\hna{H.~Nakayama}
\def\tna{T.~Natsui}
\def\yno{Y.~Noda}
\def\drn{D.R.~Noakes}
\def\yna{Y.~Nakazawa}
%
\def\jjn{J.J.~Neumeier}
\def\yoo{Y.~Ooniki}
\def\dop{D.~Opie}
\def\soh{S.~Ohkuma}
\def\mod{M.~Oda}
\def\aho{A.H.~O'Reilly}
\def\amp{A.M.~Portis}
\def\jpr{J.P.~Remeika}
\def\tmr{T.M.~Riseman}
\def\gro{G.~Roehmer}
\def\dps{D.P.~Spencer}
\def\tsu{T.~Suzuki}
\def\ysu{Y.~Suzuki}
\def\tas{T.A.~Shibata}
\def\gsh{G.~Shirane}
\def\ost{O.~Steinsvoll}
\def\mse{M.~Senba}
\def\sms{S.M.~Shapiro}
\def\cst{C.~Stassis}
\def\ces{C.E.~Stronach}
\def\hes{H.E.~Sch\"one}
\def\msu{M.~Suenaga}
\def\pws{P.W.~Schleger}
\def\dsh{D.~Shimada}
\def\asc{A.~Schenck}
%
\def\msuz{M.~Suzuki}
\def\aws{A.W.~Sleight}
\def\mas{M.A.~Subramanian}
\def\ars{A.R.~Strzelecki}
\def\bjs{B.J.~Sternlieb}
\def\cls{C.L~Seaman}
\def\bws{B.W.~Statt}
\def\pst{P.~Stamp}
\def\gsa{G.~Saito}
\def\jls{J.L.~Smith}
\def\stj{S.~Tajima}
\def\sta{S.~Takagi}
\def\mta{M.~Takigawa}
\def\jtu{J.~Tuchendler}
\def\tta{T.~Takabatake}
\def\hta{H.~Takagi}
\def\yto{Y.~Tokura}
\def\trt{T.~Thurston}
\def\ita{I.~Tanaka}
\def\lta{L.~Taillefer}
\def\yju{Y.J.~Uemura}
\def\suc{S.~Uchida}
\def\rvu{R.V.~Upasani}
%
\def\jhw{J.H.~Wernick}
\def\jpw{J.P.~Wicksted}
\def\lew{L.E.~Wenger}
\def\awi{A.~Wisiniewski}
\def\ywa{Y.~Watanabe}
\def\dlw{D.Ll.~Williams}
\def\wdw{W.D.~Wu}
\def\hhw{H.H.~Wang}
\def\jmw{J.M.~Williams}
\def\rlw{R.L.~Whetten}
\def\gxi{Gang Xiao}
\def\hya{H.~Yasuoka}
\def\tya{T.~Yamazaki}
\def\rya{R.~Yamamoto}
\def\hyo{H.~Yoshizawa}
\def\xhy{X.H.~Yu}
\def\bxy{B.X.~Yang}
\def\kya{K.~Yamada}
\def\hyam{H.~Yamochi}
\def\hzh{Hu~Zhou}

\bibitem{[1]} For general review of magnetic
and superconducting properties of HTSC systems, see,
for example,
M.A. Kastner, R.J. Birgeneau, G. Shirane, 
and Y. Endoh,
{\sl \ \ \ Magnetic, transport, and optical properties of 
monolayer copper oxides\/},
Rev. Mod. Phys. {\bf 70\/} (1988) 897.

\bibitem{[2]} R.J. Birgeneau, Y. Endoh, K. Kakurai,
Y. Hidaka, T. Murakami, M.A. Kastner, T.R. Thurston, 
G. Shirane, and K. Yamada, 
{\sl \ \ \  
Static and dynamic spin fluctuations in superconducting 
La$_{2-x}$Sr${x}$CuO$_{4}$\/},
Phys. Rev. {\bf B 39\/} (1989) 2868.

\bibitem{[3]} Y. Endoh, K. Yamada, R.J. Birgeneau, D.R. Gabbe, H.P. Jenssen,
M.A. Kastner, C.J. Peters, P.J. Picone, T.R. Thurston, J.M Tranquada,
G. Shirane, Y. Hidaka, M. Oda, Y. Enomoto, M. Suzuki,
T. Murakami,
{\sl \ \ \ 
Static and dynamic spin correlations in pure and doped La$_{2}$CuO$_{4}$\/},  
Phys. Rev. {\bf B37\/} (1988) 7443.

\bibitem{[4]}
G. Shirane, R.J. Birgeneau, Y. Endoh, P. Gehring, M.A. Kastner,
K. Kitazawa, H. Kojima, I. Tanaka, T.R. Thurston, K. Yamada,
{\sl \ \ \  Temperature Dependence of the Magnetic Excitations in 
La$_{1.85}$Sr$_{0.15}$CuO$_{4}$ ($T_{c}$ = 33 K)\/},
Phys. Rev. Lett. {\bf 63\/}
(1989) 330.

\bibitem{[5]}
G. Aeppli, T.E. Mason, S.M. Hayden, H.A. Mook, J. Kulda, 
{\sl \ \ \  Nearly Singular Magnetic Fluctuations in the 
Normal State of a High-$T_{c}$ Cuprate Superconductor\/},
Science {\bf 278\/} (1997) 1432.

\bibitem{[6]} 
H.A. Mook, P. Dai., K. Salama, D. Lee, F. Dogan, G. Aeppli, 
A.T. Boothroyd, M.E. Mostoller, 
{\sl \ \ \ 
Incommensurate One-Dimensional Fluctuations in YBa$_{2}$Cu$_{3}$O$_{6.93}$\/},
Phys. Rev. Lett. 
{\bf 77\/} (1996) 370.

\bibitem{[7]}
H.A. Mook, P. Dai, S.M. Hayden, G. Aeppli,
T.G. Perring, F. Dogan, 
{\sl \ \ \  Spin Fluctuations in YBa$_{2}$Cu$_{3}$O$_{6.6}$\/},
Nature {\bf 395\/} (1998) 580.

\bibitem{[8]}
M. Arai, T. Nishijima, Y. Endoh, T. Egami, S. Tajima, 
K. Tomimoto, Y. Shiohara, M. Takahashi, A. Garrett, S.M. Bennington,
{\sl \ \ \ Incommensurate Spin Dynamics of Underdoped Superconductor
YBa$_{2}$Cu$_{3}$O$_{6.7}$\/}, 
Phys. Rev. Lett. {\bf 83\/}
(1999) 608.

\bibitem{[9]} 
J.M. Tranquada, B.J. Sternlieb, J.D. Axe, Y. Nakamura,
S. Uchida, 
{\sl \ \ \  Evidence for Stripe Correlations of Spins and Holes in Copper Oxide
Superconductors\/},
Nature {\bf 375\/} (1995) 561.

\bibitem{[10]} J.M. Tranquada, J.D. Axe, N. Ichikawa, Y. Nakamura,
S. Uchida, B. Nachumi,
{\sl \ \ \ 
Neutron-scattering study of stripe-phase order of holes and spins in
La$_{1.48}$Nd$_{0.4}$Sr$_{0.12}$CuO$_{4}$\/},
Phys. Rev. {\bf B54\/} (1996) 7489.

\bibitem{[11]} 
J.M. Tranquada, N. Ichikawa, S. Uchida, 
{\sl \ \ \  Glassy Nature of Stripe Ordering
in La$_{1.6-x}$Nd$_{0.4}$Sr$_{x}$CuO$_{4}$\/},
Phys. Rev. {\bf B59\/} (1999) 14712.

\bibitem{[12]} 
K. Yamada, C.H. Lee, K. Kurahashi, J. Wada, S. Wakimoto, S. Ueki,
H. Kimura, Y. Endoh,
S. Hosoya, G. Shirane, R.J. Birgeneau, M. Greven, M.A. Kastner,
Y.J. Kim, 
{\sl \ \ \  Doping Dependence of the 
Spatially Modulated Dynamical Spin Correlations
and the Superconducting-transition Temperature 
in La$_{2-x}$Sr$_{x}$CuO$_{4}$\/},
Phys. Rev. {\bf B57\/} (1998) 6165.

\bibitem{[13]} T. Suzuki, T. Goto, K. Chiba, T. Shinoda, T. Fukase,
H. Kimura, K. Yamada, M. Ohashi, Y. Yamaguchi, 
{\sl \ \ \  Observation of Modulated Magnetic Long-range Order in 
La$_{1.88}$Sr$_{0.12}$CuO$_{4}$\/},
Phys. Rev. {\bf B57\/}
(1998) R3229.

\bibitem{[14]} S. Wakimoto, R.J. Birgeneau, Y.S. Lee and G. Shirane,
{\sl \ \ \ Hole concentration dependence of the magnetic moment in 
superconducting and insulating La$_{2-x}$Sr$_{x}$CuO$_{4}$\/},
Phys. Rev. {\bf B63\/} (2001) 172501.

\bibitem{[15]} B.O. Wells, Y.S. Lee, M.A. Kastner, R.J. Christianson,
R.J. Birgeneau, K. Yamada, Y. Endoh, G. Shirane, 
{\sl \ \ \  Incommensurate Spin Fluctuations  in High-Transition 
Temperature Superconductors\/},
Science {\bf 277\/} (1997) 1067.

\bibitem{[16]} Y.S. Lee, R.J. Birgeneau, M.A. Kastner, Y. Endoh, S. Wakimoto,
K. Yamada, R.W. Erwin, S.H. Lee, G. Shirane, 
{\sl \ \ \  Neutron-scattering Study of Spin-density Wave Order in the 
Superconducting State of Excess Oxygen-doped La$_{2}$CuO$_{4+y}$\/},
Phys. Rev. {\bf B60\/} (1999)
3643.

\bibitem{[17]} A. Bianconi, A.C. Castellano, M. Desantis, P. Delogu, A. Gargano,
R. Giorgi,
{\sl \ \ \ 
Localization of Cu-3d levels in the high-$T_{c}$ superconductor
YBa$_{2}$Cu$_{3}$O$_{7}$ by Cu-2p X-ray photoelectron
spectroscopy\/},
Solid State Commun. {\bf 63\/}
(1987) 1009.

\bibitem{[18]} A. Bianconi, 
{\sl \ \ \ The Instability Close to the 2D Generalized 
Wigner Polaron Crystal Density: A Possible PAiring
Mechanism Indicated by a Key Experiment\/},
Physica {\bf C235\/}-{\bf 240\/} (1994) 269. 

\bibitem{[19]}
A. Bianconi, N.L. Saini, T. Rossetti, A. Lanzara, A. Perali, 
M. Missori, H. Oyanagi, H. Yamaguchi, Y. Nishihara, D.H. Ha,
{\sl \ \ \ 
Stripe structure in the CuO$_{2}$ plane of perovskite superconductors\/},
Phys. Rev. {\bf B54\/} (1996) 12018.

\bibitem{[20]} J. Zaanen and O. Gunnarsson, 
{\sl \ \ \ Charged magnetic domain lines and the magnetism of high-$T_{c}$ 
oxides\/},
Phys. Rev. {\bf B40\/} (1989) 7391,

\bibitem{[21]} D. Poilblanc and T.M. Rice, 
{\sl \ \ \ 
Charged solitons in the Hartree-Fock approximation to the large-U
Hubbard model\/},
Phys. Rev. {\bf B39\/} (1989) 9749.

\bibitem{[22]} V.J. Emery, S.A. Kivelson and 
H.Q. Lin, 
{\sl \ \ \ Phase separation in the t-J model\/},
Phys. Rev. Lett. {\bf 64\/} (1990) 475.
\bibitem{[23]} V.J. Emery, S.A. Kivelson, O. Zachar,
{\sl \ \ \ Spin-gap proximity effect mechanism of high-temperature
superconductivity\/},
Phys. Rev. {\bf B56\/} (1997) 6120.

\bibitem{[24]} For a general review of $\mu$SR, see, for example,
A. Schenck, {\it Muon Spin Rotation Spectroscopy\/},
Adam Hilger, Bristol, 1985.

\bibitem{[25]} For a recent reviews of $\mu$SR studies in topical
subjects, see {\it Muon Science: Muons in Physics, Chemistry and
Materials\/}, Proceedings of the Fifty First Scottish 
Universities Summer School in Physics, st. Andrews, August, 1988, 
ed. by S.L. Lee, S.H. Kilcoyne, and R. Cywinski,
Inst. of Physics Publishing, Bristol, 1999.

\bibitem{[26]}
A.T.~Savici, Y.~Fudamoto, I.M.~Gat, 
M.I.~Larkin, Y.J. Uemura, K.M.~Kojima, Y.S.~Lee, M.A.~Kastner, R.J.~Birgeneau,
{\sl \ \ \  Static Magnetism in Superconducting Stage-4
La$_{2}$CuO$_{4+y}$\/}
Physica {\bf B289-290\/} (2000)
338.

\bibitem {[27]} 
Y.J Uemura,
W.J. Kossler,
X.H. Yu, J.R. Kempton, H.E. Schone, D. Opie,
C.E. Stronach, D.C. Johnston, M.S. Alvarez, D.P. Goshorn,
{\sl \ \ \  Antiferromagnetism of La$_{2}$CuO$_{4-y}$ Studied by
Muon Spin Rotation\/},
Phys. Rev. Lett. {\bf 59\/} (1987) 1045.

\bibitem{[28]}
Y.J. Uemura, \wjk, \jrk, \xhy, \hes, \dop, \ces, J.H.~Brewer, \rfk, \srk,
G.M.~Luke, \tmr, \dlw, E.J.~Ansaldo, \yen, Y.~Kudo, \kya, \dcj, M.S.~Alvarez, \dpg, \yhi, 
\mod, \yeno, \msuz, \tmu, 
{\sl \ \ \ Comparison between Muon Spin Rotation and Neutron Scattering
Studies on the 3-dimensional Magnetic Order of La$_{2}$CuO$_{4-y}$\/},
Physica {\bf C153-155\/} (1988) 769.

\bibitem{[29]}
L.P. Le, R.H.~Heffner, D.E.~MacLaughlin,
K. Kojima, G.M. Luke, B.~Nachumi, Y.J. Uemura,
J.L.~Sarrao, Z.~Fisk,
{\sl \ \ \ Magnetic Behavior in Li-doped La$_{2}$CuO$_{4}$\/},
Phys. Rev. {\bf B54\/}, (1996) 9538.

\bibitem{[30]}
D.R. Harshman, G. Aeppli, G.P. Espinosa, A.S. Cooper, J.P. Remeika,
E.J. Ansaldo, T.M. Riseman, D.L. Williams, D.R. Noakes, B. Ellman,
T.F. Rosenbaum, 
{\sl \ \ \ Freezing of Spin and Charge in La$_{2-x}$Sr$_{x}$CuO$_{4}$\/},
Phys. Rev. {\bf B38\/} (1988) 852. 

\bibitem{[31]}
B.J. Sternlieb, G.M.~Luke, Y.J. Uemura, T.M. Riseman, J.H.~Brewer,
P.M. Gehring, K. Yamada, Y. Hidaka, T. Murakami, T.R. Thurston, R.J.~Birgeneau, 
{\sl \ \ \  Muon Spin Relaxation and Neutron Scattering Studies of Magnetism
in Single-crystal La$_{1.94}$Sr$_{0.06}$CuO$_{4}$\/},
Phys. Rev. {\bf B41\/} (1990) 8866.

\bibitem{[32]}
A. Weidinger, Ch. Niedermayer, A. Golnik, R. Simon 
E. Recknagel, J.I. Budnick, B. Chamberland, C. Baines
{\sl \ \ \ Observation of Magnetic Ordering in Superconducting 
La$_{2-x}$Sr$_{x}$CuO$_{4}$ by Muon Spin Rotation\/},
Phys. Rev. Lett. {\bf 62\/} (1989) 102.

\bibitem{[33]}
F. Borsa, P. Carretta, J.H. Cho, F.C. Chou, Q. Hu, D.C. Johnston,
A. Lascialfari, D.R. Torgeson, R.J. Gooding, N.M. Salem,
K.J.E. Vos,
{\sl \ \ \ Staggered Magnetization in $^{139}$La NQR and
Mu-SR\/}, 
Phys. Rev. {\bf B52\/} (1995) 7334.

\bibitem {[34]} Ch. Niedermayer, C. Bernhard, T. Blasius, A. Golnik, 
A. Moodenbaugh, J. I. Budnick, 
{\sl \ \ \  Common Phase Diagram for 
Antiferromagnetism in La$_{2-x}$Sr$_{x}$CuO$_{4}$
and Y$_{1-x}$Ca$_{x}$Ba$_{2}$Cu$_{3}$O$_{6}$ as Seen by Muon Spin Rotation\/},
Phys. Rev. Lett. {\bf 80\/} (1998) 3843.
\bibitem{[35]} 
J.H. Brewer, E.J. Ansaldo, J.F. Carolan, A.C.D. Chaklader,W.N. Hardy,
D.R. Harshman,
M.E. Hayden, M. Ishikawa, N. Kaplan, R. Keitel, J. Kempton, R.F. Kiefl,
W.J. Kossler, 
S.R. Kreitzman, A. Kulpa, Y. Kuno, G.M. Luke, H. Miyatake, K. Nagamine,
Y. Nakazawa, N. Nishida, K. Nishiyama, S. Ohkuma, T.M. Riseman,
G. Roehmer, P. Schleger, D. Shimada, C.E. Stronach, T. Takabatake,
Y.J. Uemura, Y. Watanabe, D.L. Williams, T. Yamazaki, B. Yang.
{\sl \ \ \  Antiferromagnetism and Superconductivity in Oxygen-Defficient
YBa$_{2}$Cu$_{3}$O$_{x}$\/},
Phys. Rev. Lett. {\bf 60\/} (1988) 1073.

\bibitem{[36]} 
G.M.~Luke, \lpl, \bjs, \wdw, Y.J. Uemura, J.H.~Brewer, \tmr, 
\sis, \suc,
{\sl \ \ \  Static Magnetic Order in La$_{1.875}$Ba$_{0.125}$CuO$_{4}$\/},
Physica {\bf C185-189\/}
(1991) 1175.

\bibitem{[37]}
K. Kumagai, K. Kawano, I. Watanabe, K. Nishiyama, and K. Nagamine,
{\sl \ \ \ $\mu$SR and NMR investigations on electronic and magnetic 
state around $x=0.12$ in La$_{2-x}$Sr$_{x}$CuO$_{4}$ and 
La$_{2-x}$Ba$_{x}$CuO$_{4}$\/}, 
Hyperfine Interact. {\bf 86\/} (1994) 473.

\bibitem{[38]}
B.~Nachumi, Y.~Fudamoto, A.~Keren, 
K.M.~Kojima, M.~Larkin, G.M.~Luke, J.~Merrin, O.~Tchernyshyov,
Y.J.~Uemura, N.~Ichikawa, M.~Goto, H.~Takagi, S.~Uchida,
M.K.~Crawford, E.M.~McCarron, D.E.~MacLaughlin, R.H.~Heffner,
{\sl \ \ \  Muon Spin Relaxation Study of the Stripe Phase Order
in La$_{1.6-x}$Nd$_{0.4}$Sr$_{x}$Cu$_{2}$O$_{4}$ and Related 214 Cuprates\/}
Phys. Rev. {\bf B58\/} (1998) 8760.

\bibitem{[39]} W. Wagener, H.H. Klauss, M. Hillberg, M.A.C. de Melo,
M. Birke, F.J. Litterst, E. Schreier, B. Buchner, H. Micklitz,
{\sl \ \ \  $\mu^{+}$SR in (La$_{1.85-x}$Nd$_{x}$)Sr$_{0.15}$CuO$_{4}$\/},
Hyperfine Interact.
{\bf 105\/} (1997) 107.

\bibitem{[40]}
K.M.~Kojima, H.~Eisaki, S.~Uchida,
Y.~Fudamoto, I.M.~Gat, A.~Kinkhabwara, M.I.~Larkin,   
G.M.~Luke, Y.J. Uemura,
{\sl \ \ \  Magnetism and Superconductivity of 
High-$T_{c}$ Cuprates (La,Eu,Sr)$_{2}$CuO$_{4}$\/},
Physica {\bf B289\/} (2000) 373.

\bibitem{[41]}
H.H. Klauss, W. Wagener, M. Hillberg, W. Kopmann, H. Walf, F.J. Litterst,
M. H\"ucker, B. B\"uchner,
{\sl From Antiferromagnetic Order to Static Magnetic Stripes: The Phase
Diagram of (La,Eu)$_{2-x}$Sr$_{x}$CuO$_{4}$\/}, 
Phys. Rev. Lett. {\bf 85\/}
(2000) 4590.

\bibitem{[42]} A. Lappas, K. Prassides, F.N. Gygax, A. Schenck,
{\sl \ \ \  Magnetic and Structural Instabilities in the Stripe-phase
Region of La$_{1.875}$Ba$_{0.125-y}$Sr$_{y}$CuO$_{4}$
(0$\leq y\leq$0.1)\/},
J. Phys. Cond-Matr {\bf 12\/} (2000) 3401.

\bibitem{[43]} V. Yu. Pomjakushin, A.A. Zakharov, A.M. Balagurov, 
F.N. Gygax, A. Schenck, A. Amato, D. Herlach,
A.I. Beskrovny, V.N. Duginov, Yu.V. Obukhov, 
A.N. Ponomarev, S.N. Barilo, 
{\sl \ \ \  Microscopic Phase Separation in La$_{2}$CuO$_{4+y}$ 
Induced by the Superconducting Transition\/},
Phys. Rev. {\bf B58\/}
(1998) 12350.

\bibitem{[44]}
B. Khaykovich, Y.S. Lee, S. Wakimoto, K.J. Thomas, R. Erwin, S.-H. Lee,
M.A. Kastner, R.J. Birgeneau,
{\sl \ \ \ Enhancement of Long-Range Magnetic Order
by magnetic field in superconducting La$_{2}$CuO$_{4}$\/}
cond-mat/0112505. 

\bibitem{[45]}
L.P. Le, A. Keren, G.M. Luke, B.J. Sternlieb, W.D. Wu, Y.J. Uemura,
J.H. Brewer, T.M. Riseman, R.V. Upasani, L.Y. Chiang, W. Kang, P.M. Chaikin,
T. Csiba, G. Gruner,
{\sl \ \ \ Muon Spin Rotation/Relaxation Studies in (TMTSF)$_{2}$-X Compounds\/},
Phys. Rev. {\bf B48\/} (1993) 7284. 

\bibitem{[46]}
L.P. Le, R.H. Heffner, J.D. Thompson, G.J. Nieuwenhuys, D.E. Maclaughlin,
P.C. Canfield, B.K. Chyo, A. Amato, R. Feyerherm, 
F.N. Gygax, A. Schenck,
{\sl \ \ \ $\mu$SR studies of borocarbides\/},
Hyperfine Interact. {\bf 104\/} (1997) 49.

\bibitem{[47]} 
B. Hitti, P. Birrer, K. Fischer, F.N. Gygax, E. Lippelt,
H. Maletta, A. Schenck, M. Weber,
{\sl \ \ \ Study of La$_{2}$CuO$_{4}$ and related compounds by
$\mu$SR\/},
Hyperfine Interact. {\bf 63\/} (1990) 287.

\bibitem{[48]}
J.E. Sonier, J.H. Brewer, R.F. Kiefl, 
{\sl \ \ \ $\mu$SR Studies of the Vortex State in Type-II Superconductors\/},
Rev. Mod. Phys. {\bf 72\/} (2002) 769. 

\bibitem{[49]} 
Y.J. Uemura, G.M. Luke, B.J. Sternlieb, J.H. Brewer, J.F. Carolan, W.N. Hardy,
R. Kadono, J.R. Kempton, R.F. Kiefl, S.R. Kreitzman, P. Mulhern, T.M. Riseman,
D.Ll. Williams, B.X. Yang, S. Uchida, H. Takagi, J. Gopalakrishnan,
A.W. Sleight, M.A. Subramanian, C.L. Chien, M.Z. Cieplak, Gang Xiao, V.Y. Lee,
B.W. Statt, C.E. Stronach, W.J. Kossler, and X.H. Yu, 
{\sl \ \ \ Universal Correlations between $T_{c}$ and $n_{s}/m^{*}$ (Carrier 
Density over Effective Mass) in High-$T_{c}$ Cuprate Superconductors\/},
Phys. Rev. Lett. {\bf 62\/}, (1989) 2317.

\bibitem{[50]} 
Y.J. Uemura, L.P. Le, G.M. Luke, B.J. Sternlieb, W.D. Wu, J.H. Brewer,
T.M. Riseman, C.L. Seaman, M.B. Maple, M. Ishikawa, D.G. Hinks,
J.D. Jorgensen, G. Saito, and H. Yamochi, 
{\sl \ \ \ Basic Similarities among Cuprate, 
Bithmuthate, Organic, Chevrel-Phase 
and Heavy-Fermion Superconductors Shown by Penetration-Depth Measurements\/}, 
Phys. Rev. Lett. {\bf 66\/}, (1991) 2665.

\bibitem{[51]}
Y.J. Uemura, A. Keren, L.P. Le, G.M. Luke, W.D. Wu, Y. Kubo, T. Manako,
Y. Shimakawa, M. Subramanian, J.L. Cobb, and J.T. Markert,
{\sl \ \ \ Magnetic Field Penetration Depth in 
Tl$_{2}$Ba$_{2}$CuO$_{6+\delta}$ in the Overdoped Regime\/},
Nature (London) {\bf 364\/} (1993) 605.

\bibitem{[52]}
B. Nachumi, A. Keren, K. Kojima, M. Larkin, G.M. Luke, J. Merrin,
O. Tchernyshov, Y.J. Uemura, N. Ichikawa, M. Goto, and S. Uchida, 
{\sl \ \ \ Muon Spin Relaxation Studies of Zn-Substitution Effects
in High-$T_{c}$ Cuprate Superconductors\/}, 
Phys. Rev. Lett. {\bf 77\/}, (1996) 5421.

\bibitem{[53]}
W. Barford and J.M.F. Gunn,
{\sl \ \ \ The theory of the measurements of the London penetration depth
in uniaxial type-II superconductors by muon spin rotation\/},
Physica {\bf C156\/} (1988) 515. 

\bibitem{[54]} 
K.M. Kojima, T. Kakeshita, T. Ono, H. Eisaki, S. Uchida,
Y. Fudamoto, I. Gat, M.I. Larkin, G.M. Luke, Y.J. Uemura,
{\sl \ \ \ $\mu$SR Evidence for Coexisting Spin Stripe and 
Superconductivity Order in La$_{2-x-y}$Eu$_{y}$Sr$_{x}$CuO$_{4}$\/}, 
unpublished.

\bibitem{[55]}
Y.J. Uemura,
{\sl \ \ \ Superfluid Density, Condensation, and Phase Separation in 
High-$T_{c}$ and Other Exotic Superconductors\/},
Physica {\bf C341\/}-{\bf 348\/} (2000) 2117. 

\bibitem{[56]}
Ch. Niedermayer, C. Bernhard, U. Binninger, H. Gl\"uckler, J.L. Tallon,
E.J. Ansaldo, and J.I. Budnick, 
{\sl \ \ \ Muon Spin Rotation Study of the Correlation Between
$T_{c}$ and $n_{s}/m^{*}$ in Overdoped Tl$_{2}$Ba$_{2}$CuO$_{6+\delta}$\/},
Phys. Rev. Lett. {\bf 71\/} (1993) 1764.

\bibitem{[57]}
S.H. Pan, E.W. Hudson, K.M. Lang, H. Eisaki, S. Uchida, J.C. Davis,
{\sl \ \ \ Imaging the Effects of Individual Zinc
Impurity Atoms on Superconductivity in 
Bi$_{2}$Sr$_{2}$CaCu$_{2}$O$_{8+\delta}$\/}, 
Nature (London) {\bf 403\/} (2000) 746.

\bibitem{[58]}
Y.J.~Uemura, 
{\sl \ \ \ Bose-Einstein to BCS Crossover Picture for High-$T_{c}$ Cuprates\/}, 
Physica {\bf C282\/}-{\bf 287\/} (1997) 194.

\bibitem{[59]}
Y.J. Uemura,
{\sl \ \ \ What can we learn from comparison between cuprates and He Films ?:
Phase separation and fluctuating superconductivity\/},
Int. J. Mod. Phys. {\bf B14\/} (2000) 3003.

\bibitem{[60]}
Y.J. Uemura,
{\sl \ \ \ Microscopic phase separation in the overdoped region of
high$T_{c}$ cuprate superconductors\/},
Solid State Commun. {\bf 120\/} (2001) 347.

\bibitem{[61]}
S.H. Pan, J.P. O'Neal, R.L. Badzey, C. Chamon,
H. Ding, J.R. Engelbrecht, Z. Wang, H. Eisaki, S. Uchida, A.K. Gupta,
W.-W. Ng, E.W. Hudson, K.M. Lang,
J.C. Davis, 
{\sl \ \ \ Microscopic electronic inhomogeneity in the high-$T_{c}$
superconductor Bi$_{2}$Sr$_{2}$CaCu$_{2}$O$_{8+x}$\/},
Nature {\bf 413\/} (2001) 282.

\bibitem{[62]}
G. Agnolet, D.F. McQueeney, J.D. Reppy,
{\sl \ \ \ Kosterlitz-Thouless Transition in Helium Films\/},
Phys. Rev. {\bf B39\/} (1989) 
8934.

\bibitem{[63]}
D.J. Bishop, J.E. Berthold, J.M. Parpia, J.D. Reppy,
{\sl \ \ \ Superfluid Density of Thin $^{4}$He Films
Adsorbed in Porous Vicor Glass\/}, 
Phys. Rev. {\bf B24\/} (1981) 5047.

\bibitem{[64]}
K. Shirahama, M. Kubota, S. Ogawa, N. Wada, T. Watanabe,
{\sl \ \ \ Size-dependent Kosterlitz-Thouless Superfluid 
Transition in Thin $^{4}$He Films Adsorbed on Porous Glasses\/},
Phys. Rev. Lett. {\bf 64\/} (1990) 1541.

\bibitem{[65]}
P.A. Crowell, F.W. Van Keuls, J.D. Reppy,
{\sl \ \ \ Onset of Superfluidity 
in $^{4}$He Films Adsorbed on Disordered Substrates\/}, 
Phys. Rev. {\bf B55\/} (1997) 
12620.

\bibitem{[66]}
H. Chyo and G.A. Williams, 
{\sl \ \ \ Superfluid Phase Transition of $^{3}$He-$^{4}$He
Mixture Films Adsorbed on Alumina Powder\/}, 
J. Low Temp. Phys. {\bf 110\/} 
(1998) 533.

\bibitem{[67]}
M. Chan, N. Mulders, J. Reppy,
{\sl \ \ \ Helium in Aerogel\/},
Physics Today (August, 1996) 30.

\bibitem{[68]}
S.A. Kivelson, G. Aeppli, V.J. Emery,
{\sl \ \ \ Thermodynamics of the interplay between magnetism and 
high-temperature superconductivity\/},
Proc. Nat. Acad. Sci. {\bf 98\/} (2001) 11903.

\bibitem{[69]}
M. Matsuda, M. Fujita, K. Yamada, R. J. Birgeneau, 
M. A. Kastner, H. Hiraka, Y. Endoh, S. Wakimoto, G. Shirane, 
{\sl \ \ \ Static and dynamic spin correlations in the 
spin-glass phase of slightly doped La$_{2-x}$Sr$_{x}$CuO$_{4}$\/}, 
Phys. Rev. {\bf B62} (2000) 9148. 

\bibitem{[70]}
G.S. Boebinger, Y. Ando, A. Passner,
T. Kimura, M. Okuya, J. Shimoyama, K. Kishio,
K. Tamasaku, N. Ichikawa, S. Uchida,
{\sl \ \ \ Insulator-to-Metal Crossover in the Normal State of
La$_{2-x}$Sr$_{x}$CuO$_{4}$ Near Optimal Doping\/},
Phys. Rev. Lett. {\bf 77\/} (1996) 5417.

\bibitem{[71]}
B. Lake, G. Aeppli, K.N. Clausen, D.F. McMorrow,
K. Lefmann, N.E. Hussey, N. Mangkorntong, M. Nohara,
H. Takagi, T.E. Mason, A. Schr\"oder,
{\sl \ \ \ Spins in the Vortices of High-Temperature
Superconductor\/},
Science {\bf 291\/} (2001) 1759.

\bibitem{[72]}
E. Dagotto, T. Hotta and A. Moreo,
{\sl \ \ \ Colossal Magnetoresistant Materials: The Key Role of
Phase Separation\/},
Phys. Rep. {\bf 344\/} (2001) 1.

\bibitem{[73]}
M. Matsuda, M. fujita, K. Yamada, R.J. Birgeneau,
Y. Endoh, G. Shirane, 
{\sl \ \ \ 
Electronic Phase Separation in Lightly-doped La$_{2-x}$Sr$_{x}$CuO$_{4}$\/},
arXiv:cond-mat 011228.

\bibitem{[74]}
M.I. Larkin, A. Kinkhabwala, Y.J. Uemura, Y. Sushko, G. Saito,
{\sl \ \ \ Pressure dependence of the magnetic penetration depth
in $\kappa$-(BEDT-TTF)$_{2}$Cu(NCS)$_{2}$\/},
Phys. Rev. {\bf B64\/} (2001) 144514.

\bibitem{[75]}
Y.J. Uemura, \ake, G.M.~Luke, \lpl, \bjs, \wdw, J.H.~Brewer, \rlw, \smh,
\sol, \rbk, \fdi, \sdo, \ggr, \kho,
{\sl \ \ \ Magnetic Field Penetration Depth in K$_{3}$C$_{60}$
Measured by Muon Spin Relaxation\/}, 
Nature {\bf 352\/} 
(1991) 605.

\bibitem{[76]}
Y.J. Uemura, \ake, \lpl, G.M.~Luke, \wdw, J.S. Tsai, K. Tanigaki, 
K. Holczer, S. Donovan, R.L. Whetten,
{\sl \ \ \ 
System Dependence of the Magnetic-field 
Penetration Depth in C$_{60}$ Superconductors\/},
Physica {\bf C235\/}-{\bf 240\/} (1994) 2501.  

\bibitem{[77]}
H. Amitsuka, M. Sato, N. Metoki, M. Yokoyama, K. Kuwahara,
T. Sakakibara, H. Morimoto, 
S. Kawarazaki, Y. Miyako, J.A. Mydosh,
{\sl \ \ \ Effect of Pressure on Tiny Antiferromagnetic moment
in the Heavy-Electron Compound URu$_{2}$Si$_{2}$\/},
Phys. Rev. Lett. {\bf 83\/} (1999) 5114.

\bibitem{[78]}
K. Matsuda, Y. Kohori, T. Kohara, K. Kuwahara, H. Amitsuka,
{\sl \ \ \ Spatially Inhomogeneous Development of Antiferromagnetism
In URu$_{2}$Si$_{2}$: Evidence from $^{29}$Si NMR under Pressure\/},
Phys. Rev. Lett. {\bf 87\/} (2001) 087203.

\bibitem{[79]}
G.M. Luke, A. Keren, L.P. Le, Y.J. Uemura, W.D. Wu,
D. Bonn, L. Taillefer, J.D. Garrett, Y. Onuki,
{\sl \ \ \ Muon Spin Relaxation in Heavy Fermion Systems\/},
Hyperfine Interact. {\bf 85\/} (1994) 397.

\bibitem{[80]}
Y.J. Uemura, W.J. Kossler, X.H. Yu, H.E. Schone, J.R. Kempton, C.E. Stronach, 
S.~Barth, F.N. Gygax, B. Hitti, A. Schenck,
C.~Baines, W.F. Lankford, Y. Onuki, T. Komatsubara,
{\sl \ \ \ Coexisting Static Magnetic Order and Superconductivity in
CeCu$_{2.1}$Si$_{2}$ Found by Muon Spin Relaxation\/},
Phys. Rev. {\bf B39\/} (1989) 4726.

\bibitem{[81]}
G.M.~Luke, A. Keren, K. Kojima, L.P. Le, B.J. Sternlieb, W.D. Wu, Y.J. Uemura,
Y.~Onuki, T.~Komatsubara,
{\sl \ \ \ Competition between Magnetic Order and
Superconductivity in CeCu$_{2.2}$Si$_{2}$\/},
Phys. Rev. Lett. {\bf 73\/} (1994) 1853.

\bibitem{[82]}
P. Chandra, P. Coleman, J.A. Mydosh,
{\sl \ \ \ Pressure-induced Magnetism and Hidden Order in 
URu$_{2}$Si$_{2}$\/},
cond-mat/0110048. 

\end{thebibliography}
\end{document}